\documentclass[10pt,letterpaper]{article}
\usepackage[margin=1in]{geometry}
\usepackage{times}
\usepackage[authoryear,round]{natbib}
\usepackage[T1]{fontenc}
\usepackage{amsmath,amssymb,bm}
\usepackage{graphicx,booktabs,tabularx,array}
\usepackage{xcolor,etoolbox}
\usepackage{float}
\usepackage{placeins}
\usepackage{microtype}
\usepackage{hyperref}
\hypersetup{colorlinks=true,citecolor=blue,linkcolor=blue,urlcolor=blue,pdftitle={GeoSim: Benchmarking Multi-Field Neural Operators for Geodynamics},pdfauthor={Hailong Liu, Dinghui Yang, Mengxue Liu, Rui Qi, Meng Zhao, Yihang Tang, Jin Wen, Zhilong Fang}}
\newcommand{\dataset}{GeoSim}
\newcommand{\model}{MFGNO}
\newcommand{\pone}{GeoSim-Plume1}
\newcommand{\ptwo}{GeoSim-Plume2}

\newcommand{\fitwidth}[1]{%
  \resizebox{\ifdim\width>\linewidth\linewidth\else\width\fi}{!}{#1}}

\title{GeoSim: Benchmarking Multi-Field Neural Operators for Geodynamics}
\author{\normalsize
\begin{tabular}{c}
Hailong Liu$^{1}$, Dinghui Yang$^{1,*}$, Mengxue Liu$^{2}$, Rui Qi$^{1}$,\\
Meng Zhao$^{1}$, Yihang Tang$^{1}$, Jin Wen$^{1}$, Zhilong Fang$^{3}$\\[0.6em]
\small $^{1}$ Department of Mathematical Sciences, Tsinghua University,\\
\small Beijing 100084, China\\
\small $^{2}$ State Key Laboratory of Earthquake Dynamics, Institute of Geology,\\
\small China Earthquake Administration, Beijing 100029, China\\
\small $^{3}$ School of Resources and Environment, University of Electronic Science and Technology of China,\\
\small Chengdu, China\\
\small $^{*}$ Corresponding author: \texttt{ydh@mail.tsinghua.edu.cn}\\
\small Hailong Liu: \texttt{liuhl23@mails.tsinghua.edu.cn}
\end{tabular}}
\date{}
\begin{document}
\maketitle

\begin{abstract}
Learning reliable surrogates for geodynamic models requires benchmarks
that resolve coupled fields and temporal stability. We present \dataset, a
benchmark of high-resolution two-dimensional mantle-plume simulations
generated with I2VIS, comprising 237 trajectories across two parameter
regimes, with six coupled fields, physical timestamps, and trajectory-level
splits. \dataset\ defines short-window prediction and autonomous rollout,
and evaluates representative neural operators field by field. Across
models, we observe a pronounced scalar--velocity gap: scalar fields are
reconstructed accurately while velocity errors remain large, and
short-window accuracy does not imply stable rollout. We also evaluate
\model\ (Multi-Field Graph Neural Operator), a joint multi-field graph
baseline designed for this benchmark; on GeoSim-Plume1, under a fixed
per-field objective, it reduces horizontal and vertical velocity errors by
58.4\% and 38.5\%
relative to an independent-field
graph baseline, while an explicit field-difference input shows no
measurable benefit at this scale. \dataset\ provides a reproducible
benchmark for field-wise diagnosis and long-horizon evaluation of
geodynamic surrogates.
    
\end{abstract}

\section{Introduction}
\label{sec:introduction}

Earth's mantle dynamics govern fundamental processes including plate
tectonics, intraplate volcanism, dynamic topography, and the long-term
thermal and chemical evolution of the planet
\citep{tackley2000mantle,braun2010surface,koppers2021plumes}.
Because these processes operate at inaccessible depths and over geological
timescales, numerical geodynamic models have become a primary tool for
investigating the evolution of Earth's interior
\citep{schubert2001,gerya2019,vanZelst2022geodynamic}.
High-fidelity thermo-mechanical simulations, however, remain
computationally expensive, limiting systematic exploration of large
parameter spaces and, more critically, inverse problems that require
repeated forward or adjoint solutions to infer interior states and
rheological parameters from observations
\citep{bunge2003assimilation,li2017adjoint,ghelichkhan2024adjoint}.
Neural operators offer a promising route to amortizing these computations
by learning mappings between physical parameters and spatiotemporal
fields~\citep{li2021,kovachki2023operators}. Before such surrogates can
support reliable forward prediction and ultimately inversion, however, they
must faithfully represent the heterogeneous coupled fields produced by
realistic geodynamic models and remain reliable under temporal rollout. A surrogate may therefore reproduce
slowly varying scalar fields accurately while still failing to capture the
motion of the system, and strong short-horizon accuracy may not persist under
autonomous rollout.

\begin{figure}[t]
\centering
\includegraphics[width=\linewidth]{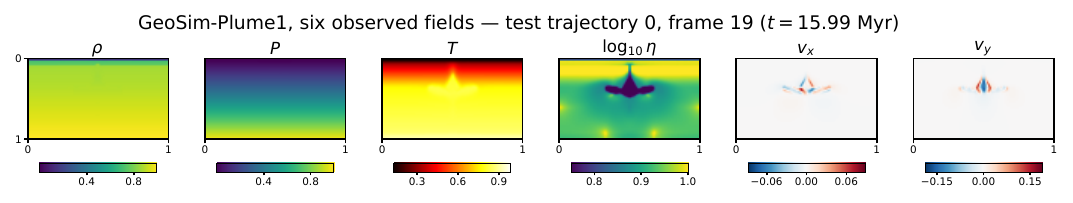}
\caption{Representative six-field state from \dataset. Density $\rho$,
pressure $P$, temperature $T$, effective viscosity $\log_{10}\eta$, and
the horizontal and vertical velocity components $(v_x,v_y)$ are
visualized at the same physical time from a representative mantle-plume
trajectory. Field values are shown in the normalized learning
representation, while the physical-domain aspect ratio is preserved. The
heterogeneous spatial structures across fields motivate evaluating
surrogate accuracy separately for each physical variable.}
\label{fig:overview}
\end{figure}

We introduce \textbf{GeoSim}, a benchmark for multi-field operator learning
in geodynamics. GeoSim contains high-resolution two-dimensional mantle-plume
simulations generated with I2VIS, with six observed fields
$(\rho,P,T,\log_{10}\eta,v_x,v_y)$ over complete physical trajectories
(Fig.~\ref{fig:overview}).
GeoSim-Plume1 varies plume temperature, while GeoSim-Plume2 jointly varies
plume temperature and lithospheric thickness. Each trajectory is treated as
an indivisible unit for data splitting, and the benchmark supports both
short-window prediction and autonomous long-horizon rollout. GeoSim
reports performance field by field, allowing differences between scalar
and velocity prediction quality to be examined explicitly.

Evaluating representative neural operators on GeoSim shows a pronounced
\emph{scalar--velocity gap}: models that predict density, pressure,
temperature, and viscosity accurately can still incur substantially larger
errors on velocity. We therefore ask whether a shared representation of
the coupled fields can reduce this gap. We design \textbf{MFGNO}
(Multi-Field Graph Neural Operator), a joint multi-field graph baseline
that jointly encodes all observed fields and performs graph-based spatial
message passing, augmented with explicit differences between neighboring
multi-field states. The design follows from the benchmark diagnosis:
instead of adding temporal machinery, MFGNO targets the spatial and
cross-field representation revealed by the field-wise analysis.

The experiments show where this baseline helps and where it fails.
Under a fixed per-field objective, MFGNO reduces the horizontal and vertical
velocity errors by $58.4\%$ and $38.5\%$, respectively, relative to an
independent-field graph model, while that model remains better on several
scalar fields. Against external baselines trained under their own
objectives, the reductions are larger, but that comparison spans objectives
and is not an isolated architecture effect. Controlled ablations show that moving from independent-field modeling to
the joint multi-field baseline is associated with a large velocity
improvement, although that comparison also changes parameter sharing and
capacity; the additional contribution of explicit field-difference inputs
is not statistically established at the current scale. Finally, autonomous rollout shows that
MFGNO's short-window velocity advantage persists over the temporal range
represented during training but does not translate into stable extrapolation
far beyond it. Thus, multi-field representation and long-horizon stability
emerge as related but distinct challenges.

Our contributions are threefold:
\begin{enumerate}
    \item \textbf{GeoSim benchmark.} We introduce a benchmark dataset and
    evaluation protocol for high-resolution mantle-plume trajectories with
    six coupled physical fields, controlled parameter variation, physical
    timestamps, and trajectory-level splits.

    \item \textbf{Field-wise diagnosis and temporal evaluation.} We identify
    a substantial scalar--velocity prediction gap and show that strong
    short-window accuracy does not imply stable autonomous rollout.

    \item \textbf{A joint multi-field graph baseline.} We design MFGNO and
    use a controlled A/B/C study to show that joint multi-field modeling is
    associated with large velocity gains under a fixed objective, while the
    additional benefit of explicit field-difference inputs is not
    established at the current scale. The joint model also reduces measured
    peak training memory and training wall time relative to the
    independent-field baseline, despite a larger parameter count.
\end{enumerate}

\section{Background and Related Work}
\label{sec:related}

\subsection{Neural Operators for PDE Surrogates}

Neural operators learn mappings between function spaces and provide a
general framework for data-driven PDE surrogates~\citep{kovachki2023operators}.
Representative approaches include Fourier Neural Operators
(FNOs)~\citep{li2021}, DeepONet~\citep{lu2021}, multipole graph neural
operators~\citep{li2020multipole}, geometry-informed
operators~\citep{li2023gino}, and convolutional neural
operators~\citep{raonic2023cno}. Graph-based learned simulators, built on
neural message passing~\citep{gilmer2017message}, include
MeshGraphNets~\citep{pfaff2021mesh} and learned physics
simulators~\citep{sanchezgonzalez2020}, as well as message-passing neural
PDE solvers~\citep{brandstetter2022}. Transformer-based operators include
GNOT~\citep{hao2023}, Transolver~\citep{wu2024}, and mesh transformers for
turbulent fluid dynamics~\citep{janny2023eagle}. Geometry-aware extensions
further adapt operator learning to irregular meshes and nonuniform
geometries~\citep{li2023geofno}. Recent work has
further scaled operator learning through PDE pretraining and flexible
spatiotemporal representations, including DPOT~\citep{hao2024dpot},
Poseidon~\citep{herde2024poseidon}, and CViT~\citep{wang2025cvit}.
Operator-learning methods have also been applied to coupled multiphysics
problems with simultaneous thermal and mechanical field
prediction~\citep{kushwaha2024multiphysics}.

Our focus is not multi-channel input per se, but the heterogeneous
field-wise difficulty that arises when coupled variables play different
roles in the underlying simulator. MFGNO uses a shared multi-field graph
representation to examine whether joint modeling can reduce this gap,
with explicit neighboring-field differences evaluated as an additional
inductive bias.

\subsection{Scientific Machine Learning Benchmarks}

A growing set of benchmarks has enabled systematic evaluation of learned
physical surrogates. PDEBench covers multiple canonical PDE families and
provides standardized datasets and baselines~\citep{takamoto2022}, while
The Well collects diverse large-scale spatiotemporal simulations under a
common interface~\citep{ohana2024well}. Open-CK explicitly evaluates
multi-physics field coupling in combustion simulations~\citep{fei2025openck},
showing that coupled-field prediction is not specific to geodynamics.
APEBench focuses on autoregressive PDE emulation and emphasizes rollout
behavior and temporal generalization~\citep{koehler2024apebench}.
WeatherBench and ClimSim provide large-scale benchmarks for data-driven
weather forecasting and hybrid physics--ML climate
emulation~\citep{rasp2020,yu2023}, respectively, and illustrate the value
of standardized splits and evaluation protocols in geoscientific
applications. Critical evaluation studies such as CoDBench and CFDONEval
compare data-driven models across continuous dynamical systems and CFD
tasks, respectively~\citep{burark2024codbench,liu2025cfdoneval}.
Long-horizon stability has also emerged as a distinct challenge for
learned PDE surrogates. PDE-Refiner systematically studies error
accumulation under recursive rollout and develops a refinement-based
strategy for improving long-term prediction
stability~\citep{lippe2023refiner}.

GeoSim targets long-term thermo-mechanical geodynamics, where temperature,
material properties, pressure, and velocity play heterogeneous roles in a
coupled numerical solver. Its evaluation emphasizes field-wise prediction
error and autonomous rollout, allowing short-window accuracy and temporal
stability to be studied separately.

\subsection{Machine Learning for Solid-Earth Dynamics}

Machine learning has already been applied to several solid-Earth simulation
problems. \citet{agarwal2021mantle} developed learned surrogates for the
evolution of two-dimensional mantle temperature fields.
More recently, \citet{agarwal2025} predicted creeping-flow velocity from
temperature while retaining a numerical solver for thermal evolution,
explicitly targeting long autoregressive mantle-convection simulation.
\citet{kong2026} studied neural operators for both forward and inverse
mantle convection, including surrogate representations of the Stokes
problem and mappings between thermal states. Related studies
of thermally driven convection have evaluated neural operators for two-
and three-dimensional Rayleigh--B\'enard systems, including long-horizon
and physical-fidelity diagnostics~\citep{markmann2026rbc}. Physics-informed
learning has also been explored for reconstructing mantle thermal
convection from partial geophysical observations~\citep{nakao2026mantle}.
Related physics--ML resources such as GlobalTomo address seismic wave
propagation
and full-waveform inversion at global scale~\citep{li2025globaltomo}, but
operate in a different physical regime from long-term thermo-mechanical
evolution.

GeoSim provides a reusable multi-field benchmark in which six coupled
geodynamic variables are modeled jointly but evaluated field by field.
This setting enables systematic study of the scalar--velocity gap and the
distinction between short-window prediction and long-horizon stability.

\section{The GeoSim Benchmark}
\label{sec:benchmark}

\subsection{Physical System and Learning Challenges}

GeoSim models the thermo-mechanical evolution of mantle plumes using the
two-dimensional I2VIS code~\citep{gerya2003,gerya2019}. Each trajectory
provides six observed physical fields,
$U(t,\mathbf{x})=(\rho,\,P,\,T,\,\log_{10}\eta,\,v_x,\,v_y)(t,\mathbf{x})$,
including density, pressure, temperature, effective viscosity, and the two
velocity components.

These channels play heterogeneous roles in the underlying simulator.
Pressure and velocity are obtained jointly from a global Stokes--continuity
solve, temperature is advanced through an implicit thermal solve, and
material-dependent quantities such as density and viscosity are reconstructed
from advected Lagrangian markers and constitutive relations. In particular,
velocity is not advanced as an independently time-integrated state. It is
recomputed from the mechanical equilibrium associated with the evolving
material configuration. GeoSim therefore tests more than multi-channel
forecasting: a learned operator must represent interactions among fields with
different dynamical and constitutive roles.

The simulations further involve strongly nonlinear visco-plastic rheology,
large spatial variations in viscosity, nonuniform spatial resolution, and
adaptive physical time stepping. Together, these properties motivate two
questions that we study explicitly: whether a model can predict all physical
fields with comparable fidelity, and whether short-window accuracy persists
under autonomous temporal rollout. Full governing equations, rheological
parameters, boundary conditions, and numerical-solver details are provided
in Appendix~\ref{app:physics}.

\subsection{Dataset Construction and Parameter Spaces}

GeoSim consists of independent mantle-plume trajectories generated on a
$1150\times660$~km domain using a $401\times309$ nonuniform Eulerian grid.
We organize them into two complementary parameter-space subsets.
Table~\ref{tab:dataset} summarizes the two settings.
\textbf{GeoSim-Plume1} varies plume temperature
over $1600$--$1800$~K using Latin-hypercube sampling.
\textbf{GeoSim-Plume2} jointly varies plume
temperature over $2000$--$2150$~K and lithospheric thickness over
$160$--$220$~km using Sobol sampling. The subset names refer to the number
of varied physical parameters. Both consist of two-dimensional simulations.

\begin{table}[t]
\centering
\small
\caption{Summary of the two GeoSim parameter-space subsets. Splits are
performed at the trajectory level.}
\label{tab:dataset}
\begin{tabular}{@{}lcc@{}}
\toprule
 & \textbf{GeoSim-Plume1} & \textbf{GeoSim-Plume2} \\
\midrule
Trajectories              & 109          & 128 \\
Train / val. / test        & 87/10/12     & 92/11/25 \\
Active parameters          & 1            & 2 \\
Plume temperature (K)      & 1600--1800   & 2000--2150 \\
Lithosphere thickness (km) & fixed        & 160--220 \\
Sampling                   & LHS          & Sobol \\
Physical fields            & 6            & 6 \\
ML resolution              & $128^2$      & $128^2$ \\
\bottomrule
\end{tabular}
\end{table}

GeoSim-Plume1 provides a controlled setting for field-wise diagnosis and
ablation, whereas GeoSim-Plume2 adds a physical degree of freedom and a
broader range of plume--lithosphere configurations. All splits are at the
trajectory level, so frames from one simulation never appear across
training and evaluation sets.

\subsection{Data Representation and Temporal Protocol}

The original fields are stored together with physical coordinates and
trajectory-specific timestamps. For learning, the nonuniform solver grid is
conservatively remapped to a common $128\times128$ representation while
preserving the field semantics and trajectory-specific physical
timestamps. Details of the coordinate reconstruction, conservative
remapping, normalization, and quality-control tests are given in
Appendix~\ref{app:data}.

An important feature of GeoSim is that physical time is not
uniformly sampled. We therefore retain the trajectory-specific timestamps
instead of replacing them with a nominal constant time step. The models in
this study are trained on the $10\to10$ window formed from the first 20
frames of each training trajectory. The 21st frame is
retained for the time axis and canonical evaluation. The released data
contain the complete trajectories. This separation allows evaluation both
within the temporal support represented during training and beyond it.

\subsection{Benchmark Tasks}

We focus on two complementary prediction tasks. In
\textbf{short-window multi-field prediction}, the model receives ten
consecutive states and predicts the following ten,
$\widehat U_{10:19}=\mathcal{G}_{\theta}(\widetilde U_{0:9},X,\mathbf{t})$,
where $X$ denotes spatial coordinates and $\mathbf{t}$ contains the physical
timestamps available to the model. Performance is reported separately for
all six fields instead of only an aggregate error, enabling direct
comparison between scalar and velocity quantities.

In \textbf{autonomous long-horizon rollout}, the first ten ground-truth
frames initialize the model, after which predicted states are recursively
used as subsequent inputs without reintroducing ground-truth state variables.
We report prediction error against physical lead time and distinguish
rollout within the temporal range represented during training from
extrapolation beyond that range.

\section{MFGNO: A Joint Multi-Field Graph Baseline}
\label{sec:mfgno}

\subsection{Motivation}

The field-wise evaluation in GeoSim motivates a simple question:
should the six physical variables be modeled as independent prediction
targets, or should their spatial evolution be represented jointly?
This distinction is particularly relevant to the mechanical fields,
especially pressure and velocity.
In the underlying simulator, pressure and velocity are obtained from a
global mechanical solve whose coefficients depend on the evolving material
and thermodynamic state. They are not advanced as isolated temporal
variables. We therefore hypothesize that exposing the coupled field history
to a shared spatial operator can improve prediction of velocity and other
mechanical fields.

\begin{figure}[t]
\centering
\includegraphics[width=\linewidth]{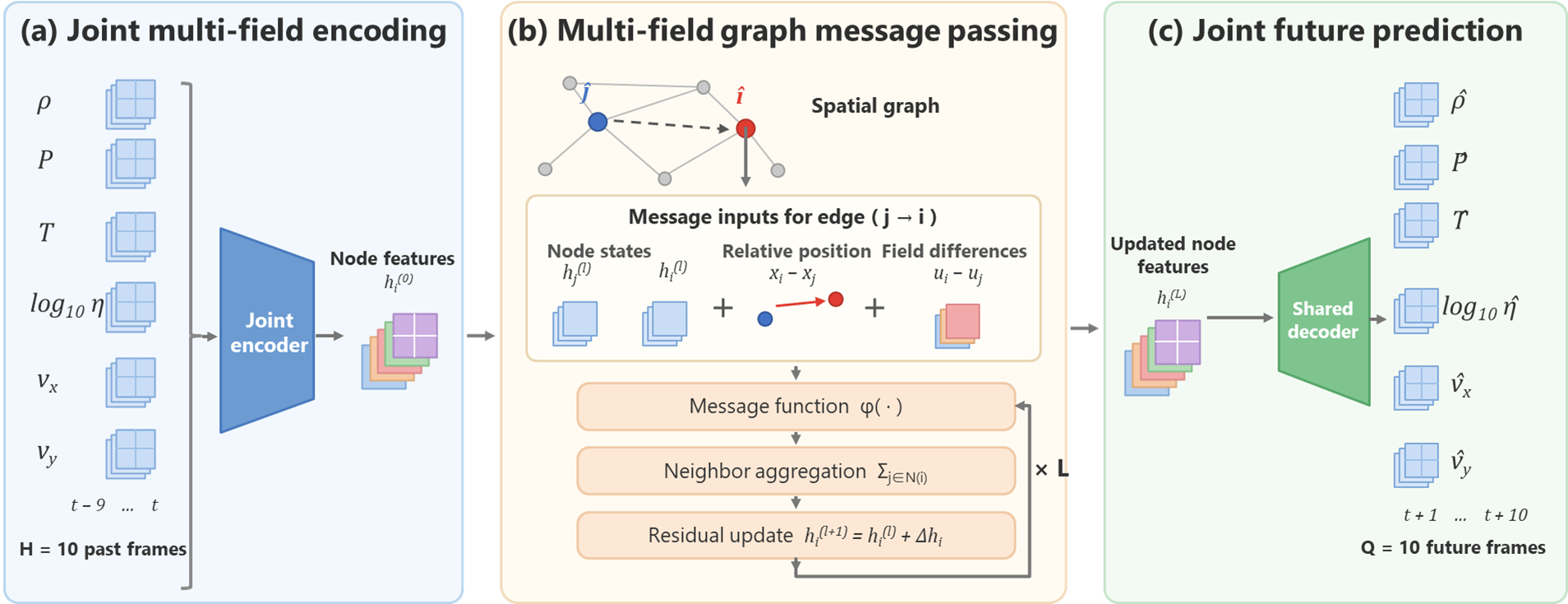}
\caption{Overview of \model. Histories of all six physical fields are
jointly encoded at each spatial node, propagated through graph-based
message passing using latent states, relative spatial offsets, and
neighboring multi-field differences, and decoded jointly to predict the
future window of all six fields.}
\label{fig:architecture}
\end{figure}

To test this hypothesis, we design \textbf{MFGNO}, a joint multi-field
graph baseline. MFGNO jointly encodes the histories of all observed
fields, propagates information over a spatial graph, and jointly decodes
their future states (Fig.~\ref{fig:architecture}). In addition, we provide
raw differences between neighboring multi-field histories to the message
function as an explicit spatial inductive bias. These two design choices are examined
separately in Sec.~\ref{sec:experiments}: the joint multi-field design is
compared with independent-field modeling, while a matched control tests
the additional value of the explicit difference pathway.

\subsection{Joint Multi-Field Representation}

Let $U_i^{\mathrm{hist}}\in\mathbb{R}^{H\times C}$
denote the history at spatial node $i$, where $H$ is the input-window
length and $C=6$ is the number of physical fields. MFGNO forms a shared
node representation from the complete field history,
$u_i=\operatorname{vec}(\widetilde U_i^{\mathrm{hist}})$,
$h_i^{0}=E_{\theta}([u_i,\mathbf{x}_i,\mathbf{c}_i])$,
where $\widetilde U$ denotes normalized fields, $\mathbf{x}_i$ denotes
spatial coordinates, and $\mathbf{c}_i$ contains the conditioning
variables used by the model.

Unlike an independent-field model, in which the predictor for field $c$
observes only the history of that field, every MFGNO node representation
has access to the histories of all six variables. This provides a common
latent state in which information from scalar and velocity fields can
interact before prediction.

\subsection{Difference-Enhanced Graph Message Passing}

Spatial interactions are modeled on a graph
$\mathcal{G}=(\mathcal{V},\mathcal{E})$ constructed from the physical
coordinates. For an edge $(j,i)$, the message at layer $\ell$ is
\begin{equation}
    m_{ij}^{\ell}
    =
    \phi_{\ell}
    \left(
    h_i^{\ell},
    h_j^{\ell},
    \mathbf{x}_i-\mathbf{x}_j,
    u_i-u_j,
    \mathbf{c}_i
    \right),
    \label{eq:mfgno_message}
\end{equation}
followed by mean aggregation and a residual node update,
\begin{align}
    \bar m_i^{\ell}
    &=
    \frac{1}{|\mathcal N(i)|}
    \sum_{j\in\mathcal N(i)}
    m_{ij}^{\ell},
    \\
    h_i^{\ell+1}
    &=
    \operatorname{IN}
    \left[
    h_i^{\ell}
    +
    \psi_{\ell}
    \left(
    h_i^{\ell},\bar m_i^{\ell},\mathbf{c}_i
    \right)
    \right],
    \label{eq:mfgno_update}
\end{align}
where $\operatorname{IN}$ denotes instance normalization.

The term $u_i-u_j$ supplies differences in the complete multi-field
history to each message layer. We treat this term as an inductive bias,
not as a calibrated physical gradient or an enforced discretization of
the governing equations. The controlled
analysis in Sec.~\ref{sec:ablation} compares the joint multi-field design
with independent-field modeling and separately tests the additional effect
of explicitly providing these differences.

\subsection{Window Prediction and Training Objective}

After $L$ message-passing layers, a shared decoder predicts the future
window for all six fields,
$\widehat U_i^{\,t+1:t+Q}=D_{\theta}(h_i^{L})$,
using the residual output parameterization implemented by the model.
For the experiments in this work, $H=Q=10$.

MFGNO and the controlled graph variants are trained with a field-wise
relative-$L^2$ objective,
\begin{equation}
    \mathcal L
    =
    \sum_{c=1}^{6}
    \lambda_c
    \,
    \mathbb E_b
    \left[
    \frac{
    \left\|
    \widehat U_b^{(c)}-U_b^{(c)}
    \right\|_2
    }{
    \left\|
    U_b^{(c)}
    \right\|_2+\epsilon
    }
    \right],
    \label{eq:mfgno_loss}
\end{equation}
where the norm is taken over the spatial and predicted temporal dimensions
and $\epsilon=10^{-8}$ is a fixed stabilizer.
We use equal field weights, $\lambda_c=1/6$, for all controlled
multi-field comparisons, so that differences between variants reflect
the information pathway and not the optimization target.

Architecture dimensions, graph construction, residual-decoder details,
conditioning variables, computational complexity, and implementation
details are provided in Appendix~\ref{app:mfgno}.

\section{Experiments}
\label{sec:experiments}

We organize the experiments around three questions:
(1) which physical fields are most difficult under the GeoSim benchmark;
(2) how a joint multi-field graph baseline compares with independent-field
modeling; and
(3) whether short-window gains persist under autonomous rollout.

\subsection{Experimental Setup}
\label{sec:setup}

We compare MFGNO with four learned baselines: FNO, GNOT, Transolver,
and an independent-field MPNN that trains a separate graph network for
each physical variable. We additionally report a persistence baseline,
which copies the last observed state across the prediction window.
MFGNO, the independent-field MPNN, and the controlled A/B/C variants
use the per-field relative-$L^2$ loss family in
Eq.~\eqref{eq:mfgno_loss}. The controlled multi-field comparisons use
equal channel weights. FNO, GNOT, and Transolver retain the
mean-squared-error objectives of their respective training pipelines,
so their rows in the main tables compare across objectives as well as
architectures. The design-attribution analysis rests on the
controlled A/B/C comparison, which holds the objective fixed.
All models use the same trajectory splits, the same $128\times128$
representation, and the same $10\!\to\!10$ prediction task. Model-specific
architectures and training budgets are fixed before evaluation. Complete configurations and compute
details are given in Appendices~\ref{app:training} and~\ref{app:cost}.

We evaluate each physical field separately using trajectory-averaged
relative $L^2$ error,
\begin{equation}
E_c =
\frac{1}{N_{\mathrm{test}}}
\sum_{n=1}^{N_{\mathrm{test}}}
\frac{1}{Q}
\sum_{q=1}^{Q}
\frac{
\left\|
\widehat U_{nq}^{(c)}-U_{nq}^{(c)}
\right\|_2
}{
\max\!\left(
\left\|U_{nq}^{(c)}\right\|_2,\epsilon_c
\right)
},
\label{eq:l2re}
\end{equation}
where $\epsilon_c$ is determined from the training split and prevents
instability for near-zero reference fields. For each subset and channel,
it is set to $10^{-6}$ times the root-mean-square of that channel over the
training file. No evaluated frame triggers this floor in the reported
results. Errors are averaged over frames within
a trajectory and then across trajectories. GeoSim-Plume1 results are
reported over three prespecified training seeds. The current GeoSim-Plume2
comparison uses one seed and is therefore treated as complementary
evidence, not a variance estimate.

\subsection{Benchmark Evaluation and the Scalar--Velocity Gap}
\label{sec:short_window}

\begin{table}[t]
\centering
\footnotesize
\caption{
Short-window field-wise relative $L^2$ error (\%) on
\pone\ under the common $10\!\to\!10$ prediction protocol.
Learned-model results are mean $\pm$ SD over three prespecified
training seeds; lower is better. Persistence is deterministic.
Best learned result in each field is shown in bold. External baselines
use their MSE objectives; graph models use the per-field relative $L^2$
objective (Sec.~\ref{sec:setup}).
}
\label{tab:main1}
\setlength{\tabcolsep}{4.5pt}
\fitwidth{\begin{tabular}{lrrrrrr}
\toprule
Model
& $\rho$
& $P$
& $T$
& $\log_{10}\eta$
& $v_x$
& $v_y$ \\
\midrule
Persistence
& 0.035
& 0.00651
& 0.509
& 0.812
& 50.1
& 45.2 \\

FNO
& $2.21{\pm}0.29$
& $2.93{\pm}0.07$
& $1.53{\pm}0.04$
& $2.50{\pm}0.08$
& $122{\pm}9$
& $115{\pm}3$ \\

GNOT
& $0.167{\pm}0.017$
& $0.318{\pm}0.017$
& $0.242{\pm}0.041$
& $0.492{\pm}0.022$
& $90.2{\pm}19.8$
& $40.6{\pm}4.7$ \\

Independent MPNN
& $\mathbf{0.0110{\pm}0.0004}$
& $\mathbf{0.00399{\pm}0.00010}$
& $\mathbf{0.0279{\pm}0.0011}$
& $\mathbf{0.0740{\pm}0.0037}$
& $11.9{\pm}1.0$
& $5.64{\pm}0.33$ \\

Transolver
& $0.0982{\pm}0.0024$
& $0.126{\pm}0.006$
& $0.332{\pm}0.013$
& $0.500{\pm}0.026$
& $61.3{\pm}2.8$
& $46.4{\pm}2.9$ \\

\model
& $0.0223{\pm}0.0007$
& $0.00508{\pm}0.00004$
& $0.0436{\pm}0.0003$
& $0.138{\pm}0.001$
& $\mathbf{4.95{\pm}0.04}$
& $\mathbf{3.47{\pm}0.05}$ \\
\bottomrule
\end{tabular}}
\end{table}

The benchmark shows a pronounced field-wise difference in prediction
quality. On GeoSim-Plume1, scalar quantities can be predicted very accurately
by relatively simple field-specific models, whereas the two velocity
components remain substantially more difficult for most evaluated operators.
For example, the independent-field MPNN obtains the lowest errors on all four
scalar channels, but its velocity errors increase to
$11.9\%$ for $v_x$ and $5.64\%$ for $v_y$.
The three external neural-operator baselines are even less accurate on
$v_x$: FNO, GNOT, and Transolver all perform worse than persistence on this
channel.

MFGNO changes this ordering on the velocity components, reaching
$E_{v_x}=4.95\pm0.04\%$ and $E_{v_y}=3.47\pm0.05\%$,
the lowest errors among the evaluated models (Tab.~\ref{tab:main1}).
Relative to the independent-field MPNN trained with the same per-field
objective, this corresponds to reductions of $58.4\%$ for $v_x$ and
$38.5\%$ for $v_y$. Against the lowest-error external baseline for each
component under its own training protocol, the reductions are $91.9\%$
and $91.5\%$. Because the external baselines use MSE objectives, we treat
this as a protocol-level comparison, not an isolated architecture effect.

This gain is not uniform across fields. The independent-field MPNN remains
better than MFGNO on density, pressure, temperature, and viscosity.
The gain is field-dependent: joint multi-field modeling trades some
scalar accuracy for a much larger gain on the velocity components.

GeoSim-Plume2 shows the same qualitative trade-off in a more difficult
two-parameter setting (full table in Appendix~\ref{app:plume2}).
MFGNO again gives the lowest velocity errors,
$82\%$ for $v_x$ and $78\%$ for $v_y$, compared with $175\%$ and $116\%$
for the lowest-error external baselines under their own protocols.
The independent-field MPNN again remains better on all four scalar fields.
These single-seed results test whether the field-wise pattern persists.
They do not characterize training variability.

The short-window results suggest that aggregate accuracy can hide
qualitatively different behavior across physical variables: models that
reconstruct the scalar state accurately need not reconstruct velocity
equally well.

\subsection{Design Study of the MFGNO Baseline}
\label{sec:ablation}

To examine the two principal design choices in MFGNO, we compare three
controlled variants:
\textbf{A}, six independent field-specific graph networks;
\textbf{B}, a joint multi-field graph network in which the explicit
difference input $u_i-u_j$ is set to zero; and
\textbf{C}, the full MFGNO.
Variants B and C have the same architecture, parameter count, training
objective, and optimization budget, and differ only in whether the explicit
multi-field difference is directly available to the message function.
Variant B still contains cross-field interaction through its joint latent
representation.

\begin{table}[t]
\centering
\small
\caption{
Controlled ablation on \pone.
A models each field independently;
B jointly models all six fields but removes the explicit neighboring-state
difference input;
C is the full \model.
All variants use the same per-field relative-$L^2$ objective with equal weights.
Values are relative $L^2$ error (\%), mean $\pm$ SD over three seeds.
}
\label{tab:ablation}
\setlength{\tabcolsep}{4.0pt}
\fitwidth{\begin{tabular}{lcccccc}
\toprule
Variant
& $\rho$
& $P$
& $T$
& $\log_{10}\eta$
& $v_x$
& $v_y$ \\
\midrule
A: Independent
& $\mathbf{0.0110{\pm}0.0004}$
& $\mathbf{0.00399{\pm}0.00010}$
& $\mathbf{0.0279{\pm}0.0011}$
& $\mathbf{0.0740{\pm}0.0037}$
& $11.9{\pm}1.0$
& $5.64{\pm}0.33$ \\

B: Joint, no diff.
& $0.0224{\pm}0.0006$
& $0.00498{\pm}0.00004$
& $0.0452{\pm}0.0012$
& $0.142{\pm}0.003$
& $4.99{\pm}0.15$
& $3.49{\pm}0.17$ \\

C: Full \model
& $0.0223{\pm}0.0007$
& $0.00508{\pm}0.00004$
& $0.0436{\pm}0.0003$
& $0.138{\pm}0.001$
& $\mathbf{4.95{\pm}0.04}$
& $\mathbf{3.47{\pm}0.05}$ \\
\bottomrule
\end{tabular}}
\end{table}

The largest change occurs between the independent-field design A and the
joint-field designs B/C. On GeoSim-Plume1, variant C reduces the $v_x$
error from $11.9\%$ to $4.95\%$ and the $v_y$
error from $5.64\%$ to $3.47\%$, while A remains
better on the four scalar channels. This contrast is consistent with the
hypothesis that joint multi-field modeling is particularly useful for the
velocity components. The comparison does not isolate a single mechanism:
A and B/C also differ in parameter sharing and have a modest
parameter-count difference.

The B--C comparison provides a cleaner test of the explicit difference
pathway. For both velocity components, the paired trajectory-level
$95\%$ trajectory-bootstrap intervals of $\Delta_c=E_{B,c}-E_{C,c}$
include zero: $\Delta_{v_x}\in[-0.0895,\,0.248]\%$ and
$\Delta_{v_y}\in[-0.0390,\,0.0858]\%$, with the full six-field
intervals in Appendix~\ref{app:ablation}.
We therefore do not find evidence that explicitly supplying
$u_i-u_j$ independently improves velocity prediction at the current
experimental scale. Small scalar differences are detectable for some
channels, but their magnitudes are much smaller and their directions are
not uniform. The evidence supports a narrower claim: the joint multi-field
design is associated with the velocity gain, whereas the additional
benefit of the explicit difference input is not established.

\paragraph{Computational cost.}
Table~\ref{tab:cost} also exposes a cost advantage of the joint design that
is not visible in the accuracy tables. In the batch-one forward-and-backward
probe, MFGNO allocates $12.9$\,GiB of peak GPU memory, compared with
$29.6$\,GiB for the independent-field MPNN, and its accumulated training
wall time is $33.2$\,h versus $52.1$\,h for the same $2000$-epoch budget
and batch size. MFGNO is not the smaller model: it has $10.55$\,M
parameters versus $9.08$\,M. The lower cost is therefore associated with
the joint implementation, which uses one shared graph network and one
invocation per ten-frame window, whereas the independent baseline runs six
field-specific networks and six invocations. Single-call inference latency
is $119$\,ms versus $213$\,ms; at the level of network invocations, the
independent baseline needs six calls per ten-frame window, so its
per-window inference cost is larger still. These are measured system-level
costs rather than a controlled compute benchmark: the memory probe is batch
one, and the independent baseline's wall time includes a resumed segment
after reaching the scheduler's $48$\,h limit.

\subsection{Long-Horizon Rollout}
\label{sec:rollout}

We next test whether the short-window velocity advantage persists under
autonomous rollout. Starting from the same ten observed frames, predictions
are recursively reused for twelve $10$-frame windows without ground-truth
reintroduction. Because timestamps are nonuniform, error is reported
against physical lead time.

\begin{figure}[t]
\centering
\includegraphics[width=\linewidth]{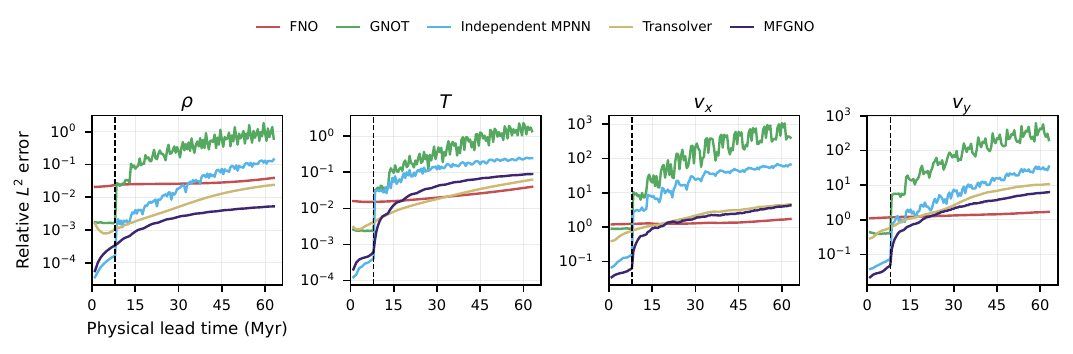}
\caption{
Autonomous rollout error as a function of physical lead time on \pone.
Relative $L^2$ error is shown for representative scalar fields $\rho$
and $T$ and both velocity components $v_x$ and $v_y$. Predictions are
recursively fed back without access to future ground-truth states. The
dashed line marks the boundary of the temporal support represented during
training. The ordinate reports relative-$L^2$ error as a fraction rather
than a percentage. Complete six-field rollout results for \pone\ and
\ptwo, including uncertainty bands, are provided in the appendix.
}
\label{fig:rollout}
\end{figure}

On \pone, the velocity advantage persists over the early rollout.
For $v_x$, \model\ reaches errors of $4.0\%$, $4.8\%$, and
$6.4\%$ at $2$, $5$, and $8$ Myr, compared with $7.7\%$, $12.7\%$, and
$16.1\%$ for the independent-field MPNN. The same holds for $v_y$
(Fig.~\ref{fig:rollout}).

This advantage does not extend to stable temporal extrapolation. At
$12$ Myr, already beyond the temporal support represented during training,
the \model\ $v_x$ error rises to approximately $57\%$. At longer horizons,
errors continue to accumulate, and \ptwo\ likewise shows no consistent
model dominance. No evaluated model satisfies the prespecified stability
criterion over the complete twelve-window rollout. Graph-model rollout
preserves the fixed-window time-conditioning convention of training. The
protocol and full results are in Appendix~\ref{app:rollout}.

\section{Discussion and Limitations}
\label{sec:discussion}

\subsection{Implications for Benchmarking Multi-Field Surrogates}

GeoSim shows that aggregate surrogate accuracy can mask strongly
field-dependent failure modes: accurate scalar reconstruction does not
guarantee accurate velocity prediction, motivating field-wise evaluation.
The gap is measured under relative $L^2$ and may reflect both field
difficulty and metric sensitivity. The present experiments do not separate
these effects from model capacity. The persistence baseline and the
consistency of the pattern across models indicate that the gap is not
specific to one architecture.

The MFGNO baseline further suggests that joint treatment of the coupled
variables is particularly beneficial for velocity prediction, consistent
with the heterogeneous structure of the thermo-mechanical system. The
experiments do not isolate a unique mechanism: the conclusion concerns
the joint multi-field baseline as a whole, while the explicit
difference pathway shows no independently established velocity benefit.

\subsection{Short-Horizon Accuracy and Temporal Stability}

Short-window accuracy and long-horizon stability emerge as distinct
challenges. Although MFGNO improves velocity prediction over short
horizons within the training temporal support, no evaluated model
satisfies the full-rollout stability criterion. This suggests that improved spatial and cross-field
representation alone is insufficient for long-horizon prediction. Future
work may require rollout-aware objectives, more faithful
continuous-time treatment, or structure-preserving constraints.

\subsection{Limitations}

GeoSim currently focuses on two-dimensional mantle-plume dynamics over a
finite range of plume temperatures and lithospheric thicknesses, and the
trajectories are computationally expensive but modest in scale relative
to large machine-learning datasets. The simulations are production runs
from a single solver and configuration. No grid-convergence study was
performed, so the benchmark describes this configuration, not a converged
limit. The present study
therefore does not establish that the observed field-wise behavior or MFGNO
advantages will transfer unchanged to other geodynamic regimes,
three-dimensional simulations, or substantially broader parameter spaces.

The current GeoSim-Plume2 comparison is based on a single training seed
and is treated as evidence that the qualitative field-wise pattern
persists. It is not a complete characterization of training variability.
Finally, the six released Eulerian fields do not capture the material and
history information the I2VIS solver evolves on Lagrangian markers.
Learned surrogates therefore approximate a partially observed system,
which may matter increasingly for long-horizon prediction.

\section{Conclusion}

We introduced \textbf{GeoSim}, a benchmark dataset and evaluation protocol
for multi-field geodynamic surrogate modeling, and used it to study
field-wise prediction difficulty and long-horizon stability. Across the
evaluated models, scalar-field accuracy does not consistently translate
into velocity accuracy, revealing a pronounced scalar--velocity gap. We also
evaluated MFGNO, a joint multi-field graph baseline designed for GeoSim.
Under a fixed per-field objective it reduces velocity error relative to an
independent-field graph baseline, while an explicit field-difference input
shows no measurable benefit at this scale. The joint design also reduces
measured peak training memory and training wall time relative to that
baseline, despite a larger parameter count. Autonomous rollout further shows
that improved short-window prediction does not by itself guarantee stable
temporal extrapolation. GeoSim thus provides a reproducible benchmark for
studying multi-field representation and long-horizon stability in learned
geodynamic surrogates.

\subsection*{Reproducibility statement}

Reproducibility details are provided throughout the appendix: the
physical configuration in Appendix A, data processing and fixed splits in
Appendix B, model and training specifications in Appendices C--D,
statistical procedures in Appendix D and ablation statistics in
Appendix H, long-horizon evaluation in Appendix I, and release provenance
in Appendix J. The GeoSim dataset is
available at
\url{https://huggingface.co/datasets/irenier/GeoSim} under CC-BY-4.0; the
per-trajectory parameter metadata is provided with the accompanying code
package. The code package provides the preprocessing, training, and
evaluation scripts and the configuration and result records used in this
study, and will be archived with a persistent identifier upon publication.

\subsection*{Ethics statement}

This work uses numerical geodynamic simulations and does not involve
human subjects, personal data, or sensitive real-world records. The GeoSim
dataset is available at
\url{https://huggingface.co/datasets/irenier/GeoSim} under CC-BY-4.0 with
documented provenance and preprocessing. The third-party I2VIS solver is
not redistributed; the dataset contains the derived simulation data, and
the accompanying code provides the authors' preprocessing and evaluation
materials.

\subsection*{AI use statement}

Generative AI tools were used for drafting and revising portions of the
manuscript, literature search, feedback on research methodology and
experimental design, assistance with portions of implementation,
interpretation and cross-checking of experimental results, and
terminological, numerical, and cross-reference consistency checks.
All AI-assisted text, code, methodological suggestions, references, and
interpretations were independently reviewed and verified by the authors
against the implementation and recorded outputs. The authors take full
responsibility for the final manuscript.

\bibliography{references}
\bibliographystyle{plainnat}
\clearpage
\appendix
\section{Physical Configuration and Observation Semantics}
\label{app:physics}

The physical interpretation of a learned predictor depends on the exact
reference problem. We therefore describe the system used to generate the
data. We distinguish what the solver \emph{implements} from what the
simulation configuration actually \emph{executes}. The simulations use
I2VIS, a marker-in-cell finite-difference code. Momentum balance is solved in both coordinate
directions,
\begin{align}
 \frac{\partial \tau_{xx}}{\partial x}
 +\frac{\partial \tau_{xy}}{\partial y}
 -\frac{\partial P}{\partial x} &= -\rho g_x,
 \label{eq:stokes-x}\\
 \frac{\partial \tau_{yy}}{\partial y}
 +\frac{\partial \tau_{xy}}{\partial x}
 -\frac{\partial P}{\partial y} &= -\rho g_y,
 \label{eq:stokes-y}
\end{align}
with $g_x=0$, $g_y=9.80665$\,m\,s$^{-2}$. Because the gravitational body
force is retained on the right-hand side against $-\nabla P$, the exported
pressure is the \emph{total} pressure: it contains the lithostatic
contribution and is not a reduced or dynamic pressure. This is the
convention used throughout this paper, and it is why we treat the pressure
field as carrying a large background component, not as a fluctuation
field.
Mass conservation is enforced through the continuity residual
\begin{equation}
 R_{\mathrm{cont}}
 =\nabla\!\cdot\!\mathbf v-d_v
 +\beta_{\mathrm{comp}}\,\frac{P-P_0}{\Delta t},
 \qquad
 \beta_{\mathrm{comp}}=\frac{1}{\rho}\frac{\partial\rho}{\partial P},
 \label{eq:continuity}
\end{equation}
where $d_v$ is a volumetric source term. The formulation is therefore
compressible, and the discrete velocity field is not divergence-free. A
hard incompressibility residual is not an appropriate diagnostic on these
fields.
Temperature is advanced with an implicit thermal solve,
\begin{equation}
 \rho C_p\frac{DT}{Dt}
 =\nabla\!\cdot(k\nabla T)+H_r+H_a+H_s,
 \label{eq:energy}
\end{equation}
with radiogenic heating, viscous dissipation $H_s$ and the configured
\emph{simplified} adiabatic term
$H_a=(\alpha T/C_p)(g_xv_x+g_yv_y)$.
Density and the thermal material properties are evaluated from the
configured material laws, including the enabled material and melt
corrections. The water/hydration subsystem accesses the thermodynamic
database, but in the simulation configuration the final density, heat
capacity and conductivity used for grid reconstruction are restored to
the material-law values. We therefore do not describe these fields as
database-derived.
Rheology is visco-plastic. Ductile deformation uses Ranalli-style
power-law parameters \citep{ranalli1995rheology} — the dry mantle
materials carry a stress exponent of $n=3.5$ and the plume material
$n=4.0$ — with a diffusion--dislocation transition and an Arrhenius factor
$\exp[(E+PV)/RT]$, so that
$\eta_{\mathrm{eff}}\propto[\eta_0\exp((E+PV)/RT)]^{1/n}
\dot\varepsilon_{II}^{(1-n)/n}$.
A Mohr--Coulomb brittle branch with strain-dependent cohesion and
friction competes with the ductile branches, the weaker mechanism
controlling the effective viscosity, and plastic yielding accumulates
strain history on the markers,
$\varepsilon^{n+1}_{\mathrm{acc}}
=\varepsilon^{n}_{\mathrm{acc}}+\Delta t\,\dot\varepsilon_{II}$.
The Peierls branch is active in the simulation configuration, applied
conditionally at low temperature and high stress ($T<1473$\,K,
$\tau_{II}>10^7$\,Pa, $\dot\varepsilon_{II}>0$), using published Peierls
data \citep{katayama2008low}.
The grid viscosity is bounded by $10^{18}\le\eta\le10^{24}$\,Pa\,s.
Two solver mechanisms are inactive in the simulation configuration and
are excluded from the present description. The alternative
mantle-rheology branch is gated on a negative rheology parameter that the
material table does not set, so it is not activated. The hard-coded
mantle phase transitions, including the 410-km transition, are commented
out of the source code. Density jumps from mantle phase transitions are
therefore not represented in this dataset.
The partial-melting routine \emph{is} evaluated during the
marker-based material update, with pressure- and water-dependent
relations for mantle peridotite \citep{katz2003new} and separate
solidus/liquidus relations for crustal materials. Whether a meaningful
melt fraction is realized in a given trajectory must be assessed from the
output data, not from the source. Melt activity is therefore not
characterized across the dataset. Melt extraction is disabled in the
simulation configuration. A
conditional hydration/material-water subsystem is enabled and contains an
antigorite-stability criterion \citep{schmidt1998experimentally}. Its
activation depends on the evolving marker state, so its realized
contribution is not quantified here.
Surface erosion and sedimentation are enabled with an erosion reference
level of $10$\,km and a sedimentation level of $30$\,km.
The exported fields are dimensional: pressure in Pa, density in
kg\,m$^{-3}$, velocity in m\,s$^{-1}$, temperature in K, and viscosity in
Pa\,s before the $\log_{10}$ transform applied at release. The domain is
$1150\times660$\,km on a $401\times309$ \emph{non-uniform} Eulerian grid
--- horizontal spacing contracts from about $4$\,km to $1$\,km through the
central plume region, and the upper $220$\,km uses $1$\,km vertical
spacing with progressive coarsening below --- with $5\times5$ Lagrangian
markers per cell over $41$ rock types. All velocity boundaries are
free-slip. The top and bottom thermal boundaries are fixed at $273$\,K
and $2113$\,K, and the lateral thermal boundaries are symmetric.
$\rho$ and $\eta$ are not free prognostic fields evolved alongside $T$:
they follow from material identity, the rheological and material-law
closures, and marker-to-grid reconstruction.

Table~\ref{tab:fields} summarizes the released channels and their grid
placement.

\begin{table}[htbp]
\centering\small
\caption{Observation dictionary. Units refer to the physical quantity
before normalization; the viscosity channel is represented as
$\log_{10}\eta$ in the learning files. The ``semantics'' column states
what the field denotes in the reference discretization; the last column
records where it is placed on the staggered grid.}
\label{tab:fields}
\begin{tabularx}{\linewidth}{@{}lllX@{}}
\toprule
Key & Symbol & Unit & Semantics and placement\\
\midrule
\texttt{ro} & $\rho$ & kg\,m$^{-3}$ & Full in-situ density from the equation of state, not an anomaly; cell center\\
\texttt{pr} & $P$ & Pa & \emph{Total} pressure including the lithostatic component; cell center\\
\texttt{tk} & $T$ & K & Absolute temperature; grid node\\
\texttt{nu} & $\log_{10}\eta$ & $\log_{10}$(Pa\,s) & Effective composite viscosity after the weakest-link combination; cell center\\
\texttt{vx} & $v_x$ & m\,s$^{-1}$ & Horizontal velocity; on the vertical cell faces\\
\texttt{vy} & $v_y$ & m\,s$^{-1}$ & Vertical velocity; on the horizontal cell faces\\
\bottomrule
\end{tabularx}
\end{table}

I2VIS uses a staggered grid with pressure, velocity, stress and strain rate
coupled through the Stokes--continuity system, and temperature advanced by
a separate thermal solver. Finite-difference and interpolation weights on
the non-uniform grid are generated with the Fornberg algorithm
\citep{fornberg1988generation}. We do not call the scheme high-order
because the simulation configuration does not enable the optional
higher-order finite-difference paths.
At each output cycle, the step is set from the stage maximum and reduced
by the marker-advection criterion. The coupled Stokes--continuity system
is solved for pressure and velocity. The step limit is recomputed from the
updated velocity. Temperature is advanced implicitly. The Lagrangian
markers are advected with a four-stage Runge--Kutta scheme. Material
properties are reconstructed on the Eulerian grid.
The Stokes--continuity system is assembled globally for $(P,v_x,v_y)$ and
handed to the Intel MKL \textbf{PARDISO} sparse \emph{direct} solver.
The solve is direct, not iterative, so there is no outer convergence
tolerance. The configured residual scales are part of the solver setup,
not a tunable accuracy budget.
The time step is adaptive and stage-dependent. It is bounded by a
stage-specific maximum, reduced according to the maximum velocity and the
permitted marker displacement,
$\Delta t\le\min(f_{\mathrm{move}}\Delta x_{\mathrm{ref}}/|v_x|_{\max},
f_{\mathrm{move}}\Delta y_{\mathrm{ref}}/|v_y|_{\max})$, and can be
reduced further by a maximum-temperature-change criterion from the
thermal solver. The startup schedule raises the cap from $1$\,yr through
$1000$\,yr to $5000$\,yr with corresponding relaxations of the
displacement factor. The internal time unit is seconds and output is
converted to years.
We did not quantify three aspects of the simulations. No grid-refinement
or resolution study was performed, so no convergence claim is made for
these runs. The discretization order is a property of the scheme, not
evidence of grid convergence. The wall-clock cost of generating the
simulations is not documented, and we do not infer it from the surrogate's
inference time. The frequency with which the conditional Peierls branch
and the partial-melting routine are activated in individual trajectories
has not been measured. Both are configured and executed as routines, but
quantifying their realized contribution requires output statistics we
have not computed. Of the $100$ Latin-hypercube designs drawn for \pone,
three are absent from the release. Their exclusion reasons were not
recorded. No other discards are known, and all $128$ \ptwo\ simulations
are present.

\section{Data Processing, Splits, and Quality}
\label{app:data}

Table~\ref{tab:dataset_inventory} summarizes the released files and split
counts.

\begin{table}[t]
\centering\small
\caption{File inventory of the GeoSim dataset. Training-pool counts precede
validation selection. Both subsets simulate two-dimensional space. The
released dataset provides complete trajectories; the trained models use a
shorter time extent (see the ``frames used in training'' row), and reported
numbers state which extent they come from. The training count is the
effective $10\to10$ target extent; the 21st frame is retained for the time
axis and canonical evaluation. Temperature and
lithosphere-thickness ranges are the simulator input intervals defined by
the sampling design, not the extrema realized by the sampled points; both
are derived from the sampling scripts and per-case inputs.}
\label{tab:dataset_inventory}
\begin{tabular}{@{}lcc@{}}
\toprule
Property & \pone & \ptwo\\
\midrule
Simulation trajectories & 109 & 128\\
Training pool / held-out test & 97 / 12 & 103 / 25\\
Final train / validation & 87 / 10 & 92 / 11\\
Frames per full train / test trajectory & 97 / 131 & 131 / 131\\
Frames used in training & 20 & 20\\
Active parameter dimension & 1 & 2\\
Temperature parameter range (K) & 1600--1800 & 2000--2150\\
Lithosphere thickness range (km) & fixed & 160--220\\
Verified sampling design & Latin hypercube, seed 42 & Sobol sequence\\
Temporal coverage (Myr) & 0--55.24 & 0--67.99\\
Stored physical channels & 6 & 6\\
Standard target resolution & $128\times128$ & $128\times128$\\
\bottomrule
\end{tabular}
\end{table}

We describe the preprocessing pipeline in execution order.
For a field stored after a transform $f_c$ and scaling $s_c^{\mathrm{data}}$,
the network value is
\[
 \widetilde U_c=f_c(U_c)/s_c^{\mathrm{data}},
\]
where the data scale is distinct from the learned decoder scale.
Only viscosity is log-transformed. The remaining fields are used with
their exported definitions, which are not uniformly raw physical
quantities.

For a cell-average interpretation, conservative remapping can be written
\[
 U'_{kl}=\sum_{i,j}W^x_{ki}W^y_{lj}U_{ij},
 \qquad
 W^x_{ki}=\frac{|I'_k\cap I_i|}{|I'_k|},
\]
with the corresponding expression in the second spatial direction.
Source intervals, boundary extrapolation, and staggered-field
collocation are specified as part of this operation. We state the
interval-overlap operator explicitly instead of describing the remapping
generically as area-aware.
\emph{Index order is part of the file format.} A field is stored as
$(N,T,n_x,n_y)$, so the \emph{second} spatial axis varies fastest:
flattening the two spatial axes maps node
$k=i\,n_y+j$ to the coordinate pair $(x_i,y_j)$. Equivalently, the node
coordinates are the \texttt{ij}-indexed mesh grid of the two coordinate
arrays --- what \texttt{torch.meshgrid(x,y,indexing="ij")} returns --- and
not the \texttt{xy}-indexed grid that \texttt{numpy.meshgrid(x,y)}
produces by default. The array indexing convention materially affects the
reported errors: using the default transposition leaves scalar channels
superficially usable while shifting their error by tens of percent, and it
reverses the two directional velocity channels, whose error can grow by an
order of magnitude. A pipeline using the default transposition can
therefore run to completion and report the wrong quantity.
Both subsets share one source grid: $401\times309$ nodes over
$1150\times660$\,km, with non-uniform spacing in both directions
(x-interval varying between $1.0$ and $4.0$\,km, y-interval between
$1.0$ and $22.1$\,km). The native solver uses a staggered grid, but the
released HDF5 fields are read out at the grid nodes and are therefore
collocated onto a common Eulerian grid. Both subsets use the same
released-grid convention, so no additional staggered-field interpolation
is required during model preprocessing. The native-grid coordinates
used for preprocessing are the verified solver coordinates, not the
uniform array the original export contained (see the quality-control
discussion in the benchmark section).
The preprocessing order is: (i) scale each field by a
\emph{max-normalization} constant, (ii) apply the conservative remapping
$U'_{kl}=\sum_{ij}W^x_{ki}W^y_{lj}U_{ij}$ with the interval-overlap
weights defined above, and (iii) attach the physical time
coordinate. Normalization therefore acts on the source grid and
remapping on normalized values. The remapping weights are non-negative
and row-normalized, so the two linear operations commute for this
particular pair. We state the order explicitly.
The weights are computed once as two one-dimensional matrices
$W^x$ ($128\times401$) and $W^y$ ($128\times309$) from the physical
interval overlaps, stored with the release, and reused for every field
and every trajectory of both subsets.
Normalization constants are computed from the pre-split training pool
(train + validation) of each subset, are fixed before model training, and
are shared by that subset's train and test files; the two subsets have
different constants and no test trajectories are used. The constants are
listed on the dataset card, and the released $128\times128$ fields are the
normalized values --- no physical-unit conversion is applied to them at
release, so downstream users who need dimensional quantities must multiply
back by the recorded constants. The native-grid files are distributed
separately in raw physical units. The logarithmic viscosity channel is normalized as
$\log_{10}\eta$ by its own maximum, not by the maximum of $\eta$; taking
the logarithm before or after averaging is \emph{not} equivalent, and no
step in this pipeline averages a viscous quantity before the transform.

\begin{figure}[h]
\centering
\includegraphics[width=\linewidth]{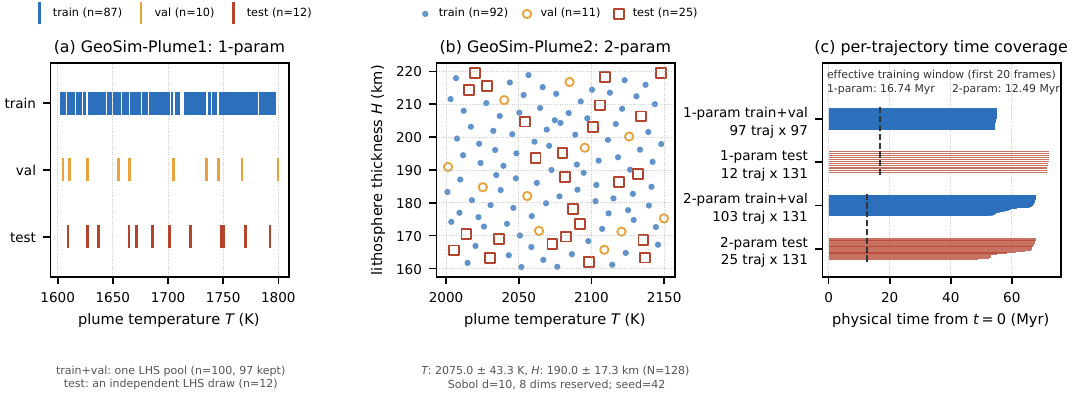}
\caption{Parameter and temporal coverage of the released trajectories.
The plot identifies partition membership and the excluded runs.}
\label{fig:coverage}
\end{figure}

A complete simulation belongs to only one of the training, validation,
and test partitions. Windows from one trajectory therefore never cross
this boundary.
The partitions are frozen by trajectory identifier, never by window, so
no trajectory contributes windows to two partitions.
\pone: the training pool is $\mathrm{T}0$--$\mathrm{T}96$, of which
$\mathrm{T}0$--$\mathrm{T}86$ are used for training and
$\mathrm{T}87$--$\mathrm{T}96$ for validation; the test set is the
separate set of simulations
$\mathrm{T\_test}0$--$\mathrm{T\_test}11$. The two identifier spaces are
independent despite overlapping numerals --- ``training index $0$'' and
``test index $0$'' are different simulations, and equality of the
integers must not be read as duplication. Train and test come from
\emph{separate simulation campaigns}, so no leakage path exists through
shared initial conditions or shared parameter draws.
\ptwo: all $128$ trajectories were split by
\texttt{default\_rng(42).permutation(128)}, the first $103$ forming the
training pool and the remaining $25$ the test set. Indices $92$--$102$
of the training pool are held out for validation. This split is
reproducible bit-for-bit and was independently re-derived from the
per-trajectory time axes, which are unique per trajectory and therefore
provide a check that does not depend on the RNG.
Leakage checks show that in both subsets the intersection of
training-pool and test identifiers is empty and their union is the
complete set. The validation block is a subset of the training pool. No
duplicate trajectories were found within any partition.
Parameter coordinates and provenance for each trajectory are recorded in
the trajectory manifest provided with the code package, and the coverage
figure (Fig.~\ref{fig:coverage}) plots the realized design, not an
idealized one.

\paragraph{Sampling design and generation.}
The sampling design uses two library calls, which makes it checkable.
\pone\ uses \texttt{scipy.stats.qmc.\allowbreak LatinHypercube}\allowbreak$(d{=}1,
\texttt{seed}{=}42)$ mapped affinely onto $[1600,1800]$~K; the training
pool and the test set are drawn at the \emph{same} seed but with different
sample counts ($100$ and $12$), so they are two independent draws, not a
split of one design.
\ptwo\ uses \texttt{scipy.stats.qmc.Sobol}$(d{=}10,
\texttt{scramble}{=}\texttt{True}, \texttt{seed}{=}42)$ with no
fast-forward: a single call draws all $128$ points. The first two
coordinates are mapped affinely onto $[2000,2150]$~K and $[160,220]$~km and
the remaining eight are unused dimensions, so the sequence is
ten-dimensional while the design it realizes is two-dimensional. The
mapping is linear and invertible, so the realized parameters inherit the
low-discrepancy property of the sequence and no post-hoc rejection or
resampling was applied --- the design is a complete Sobol$'$ point set of
size $2^7$, not a filtered subset.
Cases are named after their zero-based row in the sampled table; that
index is the trajectory index we use throughout, and it is also the
index into each split file.
Generation has five recorded stages. First, draw the design. Then create
one directory per case, copy the solver configuration into it, and
overwrite the two parameter values in the input file. Next, submit the
cases in batches. After the runs, perform the post-processing stage and
convert the binary output to HDF5. All five stages are implemented as
scripts in the accompanying code package,
and the per-case input files are retained by the authors, so every
trajectory's parameter values are recoverable without re-running anything.

Two independent checks confirm that the retained inputs match the design.
Re-running the samplers regenerates the parameter tuples of all $237$
released trajectories: the
\pone\ values agree exactly, and the \ptwo\ values agree to within the
two-decimal rounding applied when the inputs were written, at most
$0.005$~K and $0.005$~km over all $128$ cases. The values read back from
the retained inputs also reproduce the design statistics quoted in the
benchmark section. Both checks are implemented in the accompanying code
package, so the agreement is re-verified. The one dependence we cannot
remove is on the implementation of the scrambled Sobol$'$
construction: we verified reproduction under SciPy~1.10.1 and pin that
version in the reproduction environment, since a change to the scrambling
would change the design even at a fixed seed.

The continuous solver is third-party thermo-mechanical code that we neither
authored nor redistributed, so the release contains the sampling,
case-configuration, and conversion scripts together with the resulting
fields. The trajectories therefore cannot be regenerated from the release
alone. The release is sufficient to regenerate the \emph{design} and to
verify that the released fields belong to it, but not to re-run the
physics.

\paragraph{Quality control.}
A targeted scan sampled eight trajectories from each training file and
found no non-finite entries. This is a spot check, not a full-dataset
audit. Timestamps are strictly increasing within every trajectory of
every scanned file, and each trajectory carries its own time axis. In the
scanned sample, the logarithmic viscosity attains $[0.750,\,1.000]$ with
both endpoints realized, matching its declared physical bounds. We treat
this as a physical reference check of the scaling.
The original export's coordinate arrays were uniform while the solver grid
is not, and we replaced them with the verified grid metadata. The
coordinate discrepancy is zero at both domain ends and maximal in the
interior. The largest displacement is $176$\,km at the mid-plane, i.e.\
$27\%$ of the $660$\,km domain height in the $y$-direction, with the
$x$-direction worst case at $7.6\%$ of the domain width.
Because the $128\times128$ files are themselves the product of the
conservative weights, a direct
$\lVert U_{128}-W U_{401}\rVert$ comparison is identically zero up to
float32 rounding and carries no information. We instead verify the
weights against what a conservative operator must preserve: the
area-weighted integral of every field is preserved to a relative deviation
of $2\times10^{-10}$ to $6\times10^{-9}$ across the six channels --- float32
rounding, not a modeling error --- while the second-moment ratio is
$0.99999$ and the kinetic-energy ratio is $0.94$ on \pone\ and $0.85$ on
\ptwo.
Solver mesh convergence was not assessed, and no grid-refinement study was
performed. We measure conservation of nonlinear flux products such as
$v_xv_y$ under remapping. The integral
$\int v_xv_y\,\mathrm{d}A$ changes by 13--16\% between the grids across the
two subsets, consistent with the remapping preserving individual field
integrals, not nonlinear products.
A compressibility residual over the velocity fields is likewise not
reported. The solver's continuity equation is compressible by construction
(Appendix~\ref{app:physics}), so a divergence diagnostic would measure the
physical $\rho^{-1}D\rho/Dt$ term, not a discretization error. We do not
have a reference discretization against which to separate the two.

\section{MFGNO Architecture and Implementation}
\label{app:mfgno}

The reported \model\ implementation is a joint
encoder--processor--decoder over all six fields with hidden width $512$ and
depth $6$, released as class \texttt{MFGNO} in the accompanying code
package. It comprises a linear encoder mapping the $66$-dimensional node
input ($10\times6$ field values, two position coordinates, one time
coordinate, and three constant conditioning inputs, all fixed to $0.1$ in
every reported run and therefore inert) to two hidden layers of width
$1024$ and $512$ with Swish activations; six message-passing blocks of
width $512$; and a decoder producing $10\times6=60$ outputs per node.
The two position coordinates are normalized by the domain extent before
encoding. Normalization inside every message-passing block is
\emph{instance} normalization, not layer or batch normalization. The graph is built from
a radius rule whose radius is derived from the actual coordinate spacing.
It is cached and reused. Edges are not rebuilt per forward pass. The cache
is rebuilt only when the batch size changes. Each node is connected to its
immediate graph neighbors within this radius.

No physics-residual term is used in the reported experiments. The
implementation retains a per-field weight parameter that serves as the
learned decoder scale \(s_c\) in the decoder parameterization below. It
is not driven by a physics residual. \model\ and the controlled graph
variants are trained with the field-wise relative-$L^2$ objective defined
in Sec.~\ref{sec:mfgno}.

The decoder uses a residual parameterization
$\widehat{\widetilde U}_{i,q,c}
=\widetilde U_{i,\mathrm{last},c}+b_q\,s_c\,D_\theta(h_i^L)_{q,c}$,
where $s_c$ is a learned channel scale and $b_q$ is a timestamp-derived
scaling factor. The reported checkpoints use the implementation's
time-factor indexing convention. With zero-based output start index $s$,
window length $Q$, and $N$ available timestamps, for sample $b$ the
factor is
$b_q=t_{b,\min(s+Q-1+q,N-1)}-t_{b,s+Q-1}$ for $q=1,\ldots,Q$.
In the implementation, \(b_q\) is the per-sample \texttt{dt} tensor and
\(s_c\) is \texttt{physics\_weight}. Under the reported $10\to10$
evaluation with $N=21$ and $s=10$, all ten factors reduce to
$t_{20}-t_{19}$. The checkpoints therefore implement a fixed-window
time-conditioning convention. They do not use cumulative physical-time
increments across the ten predicted frames.

The three \model\ checkpoints are the seeds $0,1,2$ of the controlled
C variant in Tab.~\ref{tab:ablation}. Computational-complexity measurements
and resource use are reported in Appendix~\ref{app:cost}.

\section{Training and Evaluation Details}
\label{app:training}

\begin{table}[htbp]
\centering\scriptsize
\caption{Configuration record. Width/depth and objective are summarized from the
per-run configuration files kept alongside each checkpoint; total parameters are
counted as $\sum\mathrm{numel}$ over \texttt{model\_state\_dict} of the selected
checkpoint (for the independent-field baseline this is the \emph{sum of all six
sub-networks}). Budget is training epochs and the batch size actually used.
Seed IDs list the training seeds used in the reported short-window
and ablation results; they differ between the two subsets, which are
stated separately and not averaged.}
\label{tab:training}
\begin{tabularx}{\linewidth}{@{}lXXXXX@{}}
\toprule
Model & Width / depth & Total params & Objective & Budget / batch & Seed IDs\\
\midrule
FNO & width 48 / 4 Fourier layers, modes 24 & 10.64 M & MSE (per-frame) & 1000 ep, bs 32 & 0, 1, 2 (\pone),\newline 0 (\ptwo)\\
GNOT & hidden 128 / 3 layers, 2 experts & 2.94 M & MSE (per-frame) & 2000 ep, bs 4 & 0, 1, 2 (\pone),\newline 0 (\ptwo)\\
Independent MPNN & hidden 200 / 6 layers, 6 independent nets & 9.08 M & per-field relative $L^2$, equal weights & 2000 ep, bs 2 & 0, 1, 2 (\pone),\newline 0 (\ptwo)\\
Transolver & hidden 256 / 5 layers, 8 heads & 7.08 M & MSE (per-frame) & 1000 ep, bs 4 & 0, 1, 2 (\pone),\newline 0 (\ptwo)\\
MFGNO & hidden 512 / 6 layers, joint 6-field & 10.55 M & per-field relative $L^2$, equal weights & 2000 ep, bs 2 & 0, 1, 2 (\pone),\newline 0 (\ptwo)\\
A control & hidden 200 / 6 layers, 6 independent nets & 9.08 M & per-field relative $L^2$, equal weights & 2000 ep, bs 2 & 0, 1, 2 (\pone),\newline 0 (\ptwo)\\
B control & hidden 512 / 6 layers, zeroed difference slot & 10.55 M & per-field relative $L^2$, equal weights & 2000 ep, bs 2 & 0, 1, 2 (\pone),\newline 0 (\ptwo)\\
C control & hidden 512 / 6 layers, full joint & 10.55 M & per-field relative $L^2$, equal weights & 2000 ep, bs 2 & 0, 1, 2 (\pone),\newline 0 (\ptwo)\\
\bottomrule
\end{tabularx}
\end{table}

All reported runs use Adam with a step-decay schedule and a learning rate of
$10^{-4}$. The decay factor, step size and weight decay are not identical
across models. The external baselines were trained with their own published
schedules. These settings are recorded in the per-run configuration files
shipped with the code package, not in Tab.~\ref{tab:training}. That table
gives the
epoch budget and the batch size actually used together with the seed
identifier.
Parameters are counted from the selected
checkpoint's state dictionary, so the independent-field baseline's count
is the sum over its six sub-networks. Training uses fp32 throughout, with
no mixed precision or gradient scaling, and every reported run is
single-GPU.
Graph construction for the graph models builds a static graph and reuses
it. The edge set is not recomputed per forward pass, and the radius is
derived from the actual coordinate spacing. The graph is rebuilt only when
the batch size changes. The
relative cost of rebuilding versus reusing that graph is measured
separately and reported in Appendix~\ref{app:cost}.
Gradient scope: the models are trained on the $10\to10$ window with the
pushforward depth fixed at one window; we do not back-propagate through
multiple chained windows, so no long-horizon gradient signal is used
during training.
Validation uses mean-squared error as the checkpoint-selection criterion.
The training objective for the graph models is the per-field relative
$L^2$ loss. There is no early stopping based on test data and no
post-hoc time rescaling. No hyperparameter search was performed. Each
model is trained once per seed at a budget chosen before the runs, so no
model is tuned to its optimum.
The controlled ablation uses a \emph{single} equal-field objective for
A, B, and C; the velocity-weighted variant of the loss family is
implemented in the code package but not trained in this study.
The fixed $\epsilon=10^{-8}$ used in the training objective
(Eq.~\eqref{eq:mfgno_loss}) is distinct from the channel-specific
evaluation stabilizer $\epsilon_c$ in Eq.~\eqref{eq:l2re}.
The graph checkpoints use the time-conditioning convention described in
Appendix~\ref{app:mfgno}. In the canonical 21-frame evaluation,
the decoder factors are constant across the ten outputs of a sample,
$b_q=t_{b,20}-t_{b,19}$, and may differ between samples.
Long-horizon evaluation repeats this learned window operator without
replacing its factor by a different cumulative-increment formula.

Each result record used in this study stores the dataset checksum,
configuration, model-source version, seed, checkpoint, and metric
definition. Results obtained with different time normalization, loss
weights, prediction windows, or aggregation rules are treated as separate
experimental records.

\paragraph{Statistical reporting.}
Frame-level errors are aggregated within a trajectory before averaging
across trajectories. The seed mean and standard deviation summarize
training variability. For paired trajectory resampling, all frames of a
trajectory are retained as one unit, so frames are never treated as
independent. We report two uncertainty summaries that are not
interchangeable. The \emph{seed standard deviation} is taken over the
three training seeds and describes training variability. It is a standard
deviation, not an interval. The \emph{trajectory interval} is
a percentile bootstrap over test trajectories with $10^{4}$ resamples and
a fixed resampling index set shared by all models, so that comparisons
between models use identical resamples. The reported interval conditions
on the trained checkpoints and therefore does not include training
variability, so it is narrower than a seed-to-seed interval would be.
Paired comparisons of the form $E_B-E_C$ subtract per trajectory
\emph{before} resampling, which removes the shared trajectory effect and
gives a tighter interval than differencing two separately bootstrapped
means.
Near-zero norms: when a channel's reference norm over a frame falls below
the fitted threshold $\epsilon_c$, Eq.~\eqref{eq:l2re} substitutes
$\epsilon_c$ for that norm so that the ratio remains finite, and the
frame is counted as a trigger; trigger counts are reported alongside the
aggregate. This substitution is a stabilizer. It does not imply that the
error is small: for such a frame the relative error is numerically
ill-conditioned, and the count states how often this occurs. No channel
triggered in the results reported here.
All models are evaluated on the full test set. We do not test for
equivalence between models. Differences are reported as ``not established
at this scale''.

\paragraph{Additional error definitions.}
The area-weighted frame RMSE is
\[
 r_{nqc}=\left[\sum_i a_i
 (\widehat U_{nqic}-U_{nqic})^2\right]^{1/2}.
\]
A channel's persistence skill can be computed as
\[
 S_c=1-\frac{\sum_{n,q,i}a_i
 (\widehat U_{nqic}-U_{nqic})^2}
 {\sum_{n,q,i}a_i
 (U_{n,\mathrm{last},i,c}-U_{nqic})^2}.
\]
For short windows, the last observation is the initial true history's
final frame. For rollout, the baseline holds that same initial
observation, without accessing subsequent truth.
A zero denominator yields an undefined skill. The score is reported as
undefined.
These complementary metrics are not interchangeable with the
trajectory-average relative error.

The main tables can be reproduced from machine-readable records. Each
record states the trajectory count and frame count it was computed over.
The records are:
\begin{itemize}\itemsep2pt
\item per-run evaluation records, one file per run, holding the six
per-field $E_c$, the six per-field RMSE values, the per-field
$\epsilon$-trigger counts, the trajectory and frame counts, the selected
checkpoint's epoch and validation loss, and the test file actually read;
\item a consolidated ablation table over the three seeds with the same
fields, for variants A, B and C;
\item a consolidated external-baseline table for FNO, GNOT and
Transolver under the identical evaluation protocol;
\item a per-trajectory record holding $e[n,q,c]$ (per frame, per field)
and its trajectory mean $E[n,c]$, saved for the ablation runs whose paired
intervals are quoted below, so that those intervals can be recomputed
without re-running inference;
\item the persistence baseline record, computed with the anchor frame
declared in the protocol, not chosen per model;
\item the spatial diagnostic records: the transect profile amplitudes
and peak displacements for every model, and the declared failure-case
selection.
\end{itemize}
The relative $L^2$ error follows Eq.~\eqref{eq:l2re}; RMSE is reported
in the normalized field units, not in physical units, because the
$128\times128$ fields used here are normalized.
We do not report structural similarity (SSIM) because its settings were
not frozen for this protocol. Reporting it alongside Eq.~\eqref{eq:l2re}
would compare two different conventions.
The skill score $1-\mathrm{SSE}_{\text{model}}/\mathrm{SSE}_{\text{baseline}}$
is defined in the protocol; when the baseline's error for a channel and
frame is exactly zero the score is recorded as undefined, not assigned a
floor value.
Failure cases use a prespecified rule: the trajectory on which most
models have their largest velocity error. The selected identifier is
recorded with the figure.

\section{Computational Cost and Scalability}
\label{app:cost}

Table~\ref{tab:cost} reports the measured accuracy--cost record.

\begin{table}[htbp]
\centering\small
\caption{Measured accuracy--cost record. Parameters are counted from checkpoints;
training hours are the \emph{accumulated job wall time} from \texttt{sacct},
including segments killed by the original 48\,h submission limit and any
resumed runs (the released training script raises this limit to 96\,h to
avoid segmentation), so they bound cost, not pure compute. Single-call inference latency is batch
one, fp32, on the same GPU partition (GNOT's graph remains on the CPU); it is the measured latency of one model
invocation and should not be interpreted as end-to-end latency for an
identical ten-frame prediction task, because the number of invocations
required differs across model families (one per window for the joint graph
models and GNOT, six for the independent-field baseline, and ten for FNO and
Transolver). $E_{v_x}/E_{v_y}$ are the seed-mean errors from the common
$10\!\to\!10$ protocol reported in Tab.~\ref{tab:main1} in percent, so the
two tables can be read against each other. Peak memory is a batch-one probe: forward \emph{and backward} for the
graph models, forward only for the external baselines, whose training code was
not instrumented --- the five rows are comparable within those two groups and
not across them.}
\label{tab:cost}
\fitwidth{\begin{tabular}{@{}lccccc@{}}
\toprule
Model & Params (M) & Peak GiB & Train hours & Single-call ms & $E_{v_x}/E_{v_y}$ (\%)\\
\midrule
FNO & 10.64 & 0.15 & 2.39 & 2.8 & 122 / 115\\
GNOT & 2.94 & 1.85 & 4.32 & 16.8 & 90.2 / 40.6\\
Independent MPNN & 9.08 & 29.57 & 52.14 & 213.3 & 11.9 / 5.64\\
Transolver & 7.08 & 1.25 & 17.28 & 28.6 & 61.3 / 46.4\\
MFGNO & 10.55 & 12.88 & 33.24 & 119.4 & 4.95 / 3.47\\
\bottomrule
\end{tabular}}
\end{table}

\paragraph{Hardware and software.}
All measurements were taken on a single GPU node of the same cluster
partition used for training, with one process per model and no other job
sharing the device. The recorded environment is a single NVIDIA A800-SXM4-80GB (79.33~GiB of
device memory, compute capability 8.0), CUDA 12.1, PyTorch 2.4.1 and
Python 3.8.20. A short probe records these values in a machine-readable
environment file included with the code, so the environment a measurement
was taken in can be checked.
Every model runs in float32.

\paragraph{Timing scope.}
Inference latency is a single network forward pass at batch one, after
warm-up and with an explicit device synchronization around each
measurement; repeated measurements are used to estimate timing
variability, and the raw timing statistics are included in the code
package. Graph construction is measured
in two separate modes because they differ by an order of magnitude: a full
rebuild including neighbor search, and a cached update that replaces node
features on an existing edge index. The measured values are
$6.21\pm0.07$\,ms for a full rebuild and $0.60\pm0.07$\,ms for a cached
update --- a factor of ten --- for a graph of $16{,}384$ nodes and
$129{,}540$ directed edges, corresponding to $7.9$ neighbors per node. The reported peak memory is
\texttt{torch.cuda.max\_memory\_allocated} over one training step
approximated as \emph{graph construction plus forward plus backward}. It
therefore includes edge-message activations and gradients but
\emph{excludes} optimizer state, since no optimizer step is taken; runs
that would exceed device memory are not silently truncated.
Training hours are accumulated job wall time, summed over every segment a
run occupied including segments terminated by the scheduler's time limit,
and taken from the scheduler's own accounting, not from log timestamps.
Warm-up and repetition counts differ by measurement --- graph rebuild and
external-baseline inference use two warm-up and five timed iterations, while
the cached graph update and the graph models' inference use three and ten ---
because the rebuild is expensive enough that more repetitions would dominate
the measurement budget without reducing its variance.

\paragraph{Comparability.}
The five rows share one node, one precision, one batch definition and one
timing boundary, but the amount of predicted output per call differs
across model families, so single-call latency is a per-call cost and
not an end-to-end cost for an identical ten-frame prediction task. For
GNOT, the graph structure remains on the host because the installed DGL
build is CPU-only, so its latency includes CPU-side graph operations; the
PyG graph models keep the graph on the device. The models and their inputs
run on the GPU exactly as they do during training, and the reported latency
reflects this difference.
We do not report a solver speed-up. The solver's cost is not measured
here, and the tasks are not matched: the network predicts ten frames in
one call while the solver advances its own time step under a stability
constraint,
so any ratio between the two numbers would compare quantities that do not
correspond.

\section{Field Predictions and Spatial Diagnostic}

The spatial diagnostic uses a prespecified transect. The transect is the
vertical line at $x^\ast=0.5$ of the normalized domain, i.e.\ the physical
mid-plane of the $1150\times660$\,km box. It falls on the interface
between grid columns 63 and 64, so the two adjacent columns are averaged.
The displayed time is the last frame of the $10\to10$ output
window (frame index 9, the largest lead time in the protocol) and the example
trajectory is fixed a priori as trajectory 0 of the frozen test set.
Amplitude and displacement are reported separately, because a relative $L^2$
error cannot distinguish a profile of the wrong magnitude from a correct
profile at the wrong height.

\begin{figure}[H]
\centering
\includegraphics[width=\linewidth]{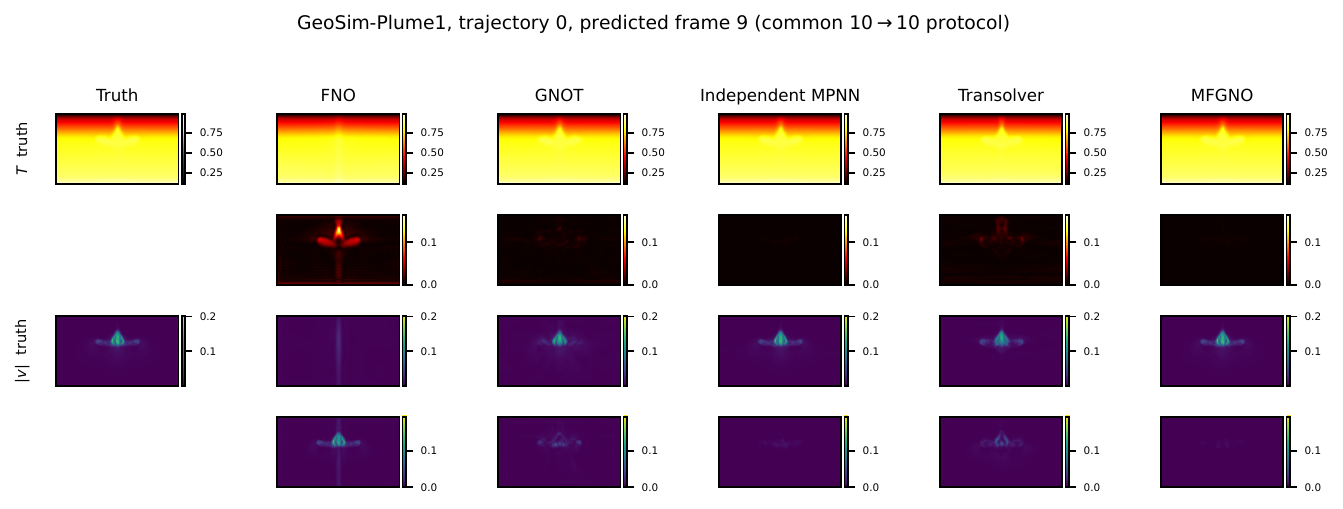}
\caption{Field-wise predictions for \pone\ trajectory 0 at the
final frame of the common $10\!\to\!10$ prediction window. All models
use the same trajectory, physical time, coordinate convention, and
field-wise color scales. Values are shown in normalized units.}
\label{fig:fields}
\end{figure}

\begin{figure}[H]
\centering
\includegraphics[width=\linewidth]{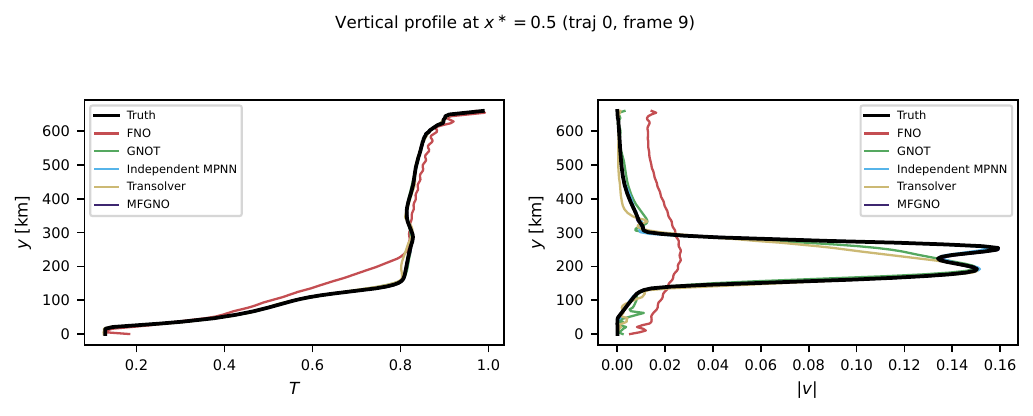}
\vspace{4pt}
\includegraphics[width=\linewidth]{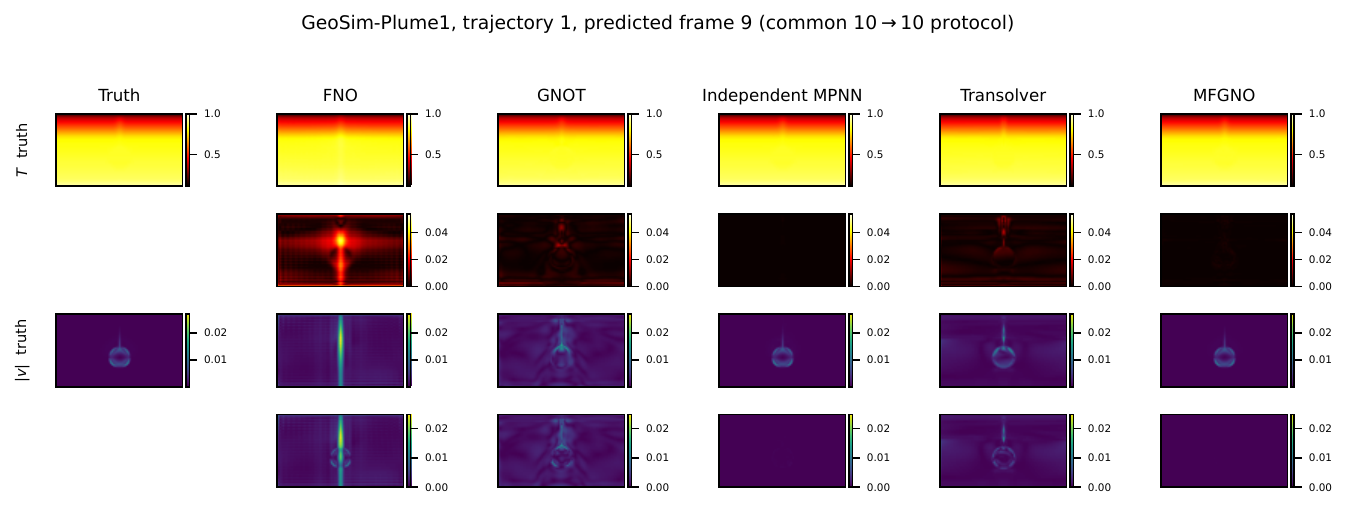}
\caption{The declared spatial diagnostic, separated from the field panels.
Both panels of the upper row are the same vertical transect at the physical
mid-plane $x^\ast=0.5$, averaged over the two grid columns that straddle it
and shown for the illustrative trajectory and frame of
Fig.~\ref{fig:fields}: temperature on the left, speed $|v|$ on the right.
Amplitude and peak displacement are read from this quantity because the
relative $L^2$ error cannot distinguish a profile of the wrong magnitude
from a correct profile at the wrong height. The aggregate over all twelve
test trajectories is quoted in the text. Profile values are shown in
normalized units. The lower panel shows the trajectory selected by the
prespecified rule, and it is
distinct from the illustrative trajectory above.}
\label{fig:profile}
\end{figure}

Fig.~\ref{fig:profile} shows the declared transect and the selected
failure case. For the temperature profile every model is essentially exact
in both respects (amplitude error within $1.5\%$, peak displacement at most
$5.2$\,km). Velocity profiles exhibit substantially larger model
differences. FNO underestimates the velocity amplitude by a median of
$82\%$, with a maximum overshoot of $230\%$ on one trajectory.
GNOT and Transolver have smaller median amplitude errors ($-7.9\%$ and
$-0.9\%$) but larger peak-position errors, with median displacements of
$7.8$\,km and $20.8$\,km and maximum displacements of $99$\,km and
$104$\,km. \model\ and the independent-field MPNN (variant A) show smaller
median amplitude errors ($+0.1\%$ and $-0.2\%$) and peak displacements
($0.0$\,km in the median, at most $5.2$\,km on any trajectory). The
transect is a single line, so the diagnostic describes structure along
one cut. The per-trajectory records behind these numbers are
included with the figure in the code package. Peak displacements are
quantized by the $128\times128$ grid spacing (about $5.2$\,km), so the
reported $5.2$, $99$, and $104$\,km correspond to 1, 19, and 20 grid
intervals.

\section{GeoSim-Plume2 Short-Window Results}
\label{app:plume2}

Table~\ref{tab:main2} reports the \ptwo\ field-wise results discussed in
Sec.~\ref{sec:short_window}.

\begin{table}[H]
\centering
\small
\caption{\ptwo\ short-window field-wise relative $L^2$ error (\%) under the
common $10\!\to\!10$ prediction protocol; lower is better. Rows are
single-seed runs and carry no seed standard deviation (FNO, GNOT,
Independent MPNN, Transolver, \model). Persistence holds the last observed
input frame (frame index 9) constant over the prediction window, and by
protocol carries no seed variance. Best learned result in each field is
shown in bold. External baselines use their MSE objectives; graph models
use the per-field relative $L^2$ objective (Sec.~\ref{sec:setup}).}
\label{tab:main2}
\setlength{\tabcolsep}{4.5pt}
\begin{tabular}{lcccccc}
\toprule
Model
& $\rho$
& $P$
& $T$
& $\log_{10}\eta$
& $v_x$
& $v_y$ \\
\midrule
Persistence
& 1.03
& 0.188
& 2.58
& 2.18
& 213
& 221 \\

FNO
& 2.68
& 3.01
& 4.16
& 2.79
& 356
& 121 \\

GNOT
& 0.514
& 0.389
& 1.01
& 1.04
& 276
& 133 \\

Independent MPNN
& $\mathbf{0.432}$
& $\mathbf{0.131}$
& $\mathbf{0.819}$
& $\mathbf{0.938}$
& 109
& 94.0 \\

Transolver
& 0.579
& 0.244
& 1.02
& 1.10
& 175
& 116 \\

\model
& 0.565
& 0.151
& 0.973
& 1.08
& $\mathbf{82.0}$
& $\mathbf{78.0}$ \\
\bottomrule
\end{tabular}
\end{table}

\section{Controlled Ablation Statistics}
\label{app:ablation}

Table~\ref{tab:ablation_stats} reports the paired B--C differences and the
full six-field trajectory-level intervals whose velocity components are
quoted in Sec.~\ref{sec:ablation}.

\begin{table}[H]
\centering
\small
\caption{Paired B--C differences and $95\%$ paired trajectory-bootstrap
intervals on \pone, in relative $L^2$ error (\%). $E_B-E_C$ reports the
mean $\pm$ SD over the three paired training seeds, while the $95\%$
intervals are trajectory-level paired bootstrap intervals ($10^{4}$
resamples, one index set shared across models) conditioning on the trained
checkpoints.}
\label{tab:ablation_stats}
\setlength{\tabcolsep}{4.5pt}
\begin{tabular}{lccc}
\toprule
 & $\rho$ & $P$ & $T$\\
\midrule
$E_B-E_C$ & $(+7.21{\pm}1.48){\times}10^{-5}$ & $(-9.42{\pm}1.88){\times}10^{-5}$ & $(1.62{\pm}1.40){\times}10^{-3}$\\
95\% traj.~boot. & $[1.16{\times}10^{-5},\,1.41{\times}10^{-4}]$ & $[-1.29{\times}10^{-4},\,-6.11{\times}10^{-5}]$ & $[9.70{\times}10^{-4},\,2.32{\times}10^{-3}]$\\
\bottomrule
\end{tabular}

\vspace{4pt}

\begin{tabular}{lccc}
\toprule
 & $\log_{10}\eta$ & $v_x$ & $v_y$\\
\midrule
$E_B-E_C$ & $(3.52{\pm}2.04){\times}10^{-3}$ & $(4.53{\pm}4.06){\times}10^{-2}$ & $(1.64{\pm}6.20){\times}10^{-2}$\\
95\% traj.~boot. & $[1.99{\times}10^{-3},\,4.99{\times}10^{-3}]$ & $[-0.0895,\,0.248]$ & $[-0.0390,\,0.0858]$\\
\bottomrule
\end{tabular}
\end{table}

\section{Long-Horizon Rollout Protocol}
\label{app:rollout}

\paragraph{Protocol.}
Each rollout starts from the same ten observed frames used by the
short-window task. Every later window is built from the model's own
previous outputs, and no ground-truth state is reintroduced between calls.
The window stride equals the output length ($10$ frames), so successive
windows abut without overlap and no frame is predicted twice. No temporal
resampling is performed --- the non-uniform stored timestamps are used as
given, and each output frame is scored against the true frame with the same
index. Because the stored snapshot intervals are non-uniform and differ
between trajectories, endpoints are reported as \emph{fixed lead times}
measured from the last observed input frame, obtained by interpolating the
per-frame error curve within each trajectory's own support; nothing is
extrapolated beyond a trajectory's end. The effective training temporal
support in lead-time terms is bounded by $t_{19}-t_9$, the last target
frame measured from the last input frame of the first window. On \pone\
this boundary is $8.0$\,Myr on every trajectory. The $2$ and $5$\,Myr
endpoints lie inside the support for all twelve test trajectories, the
$8$\,Myr endpoint is at the boundary, and the $12$\,Myr endpoint lies
outside for all of them. On \ptwo\ the boundary varies per trajectory.
The $2$ and $5$\,Myr endpoints lie inside for most trajectories, the
$8$\,Myr endpoint is outside for most, and the $12$\,Myr endpoint lies
outside for all $25$.

\paragraph{Number of model calls.}
The number of calls differs by model family even at equal frame count:
the graph models and GNOT emit ten frames per call, whereas FNO and
Transolver emit one frame per call and need ten calls per
window. Within the graph models the joint model makes one network call per
window, while the independent-field baseline calls six sub-networks, one
per field, so its call count is six times larger at the same frame count.

\paragraph{Time conditioning and rollout compatibility.}
The graph models' time conditioning is a fixed-horizon operator, not a
continuous-time one: their decoder builds its time factors from indices
that, in the canonical twenty-one-frame evaluation, all resolve to the
last interval $t_{20}-t_{19}$. Supplying the full-length time axis changes that
operator immediately, already at the first predicted window, so the rollout
reported here runs the graph models through a compatibility layer that
reproduces the trained window operator at each block while keeping absolute
time conditioning at the window start. The checkpoints therefore use a
fixed-horizon operator without changing weights or architecture. The
layer does not implement cumulative physical-time increments, and
a model that did would have to be retrained and evaluated separately.
FNO and Transolver are per-frame autoregressive and are unaffected by
this layer. GNOT emits a ten-frame window per call and does not
use the graph-model time-conditioning rule.

\begin{figure}[H]
\centering
\includegraphics[width=\linewidth]{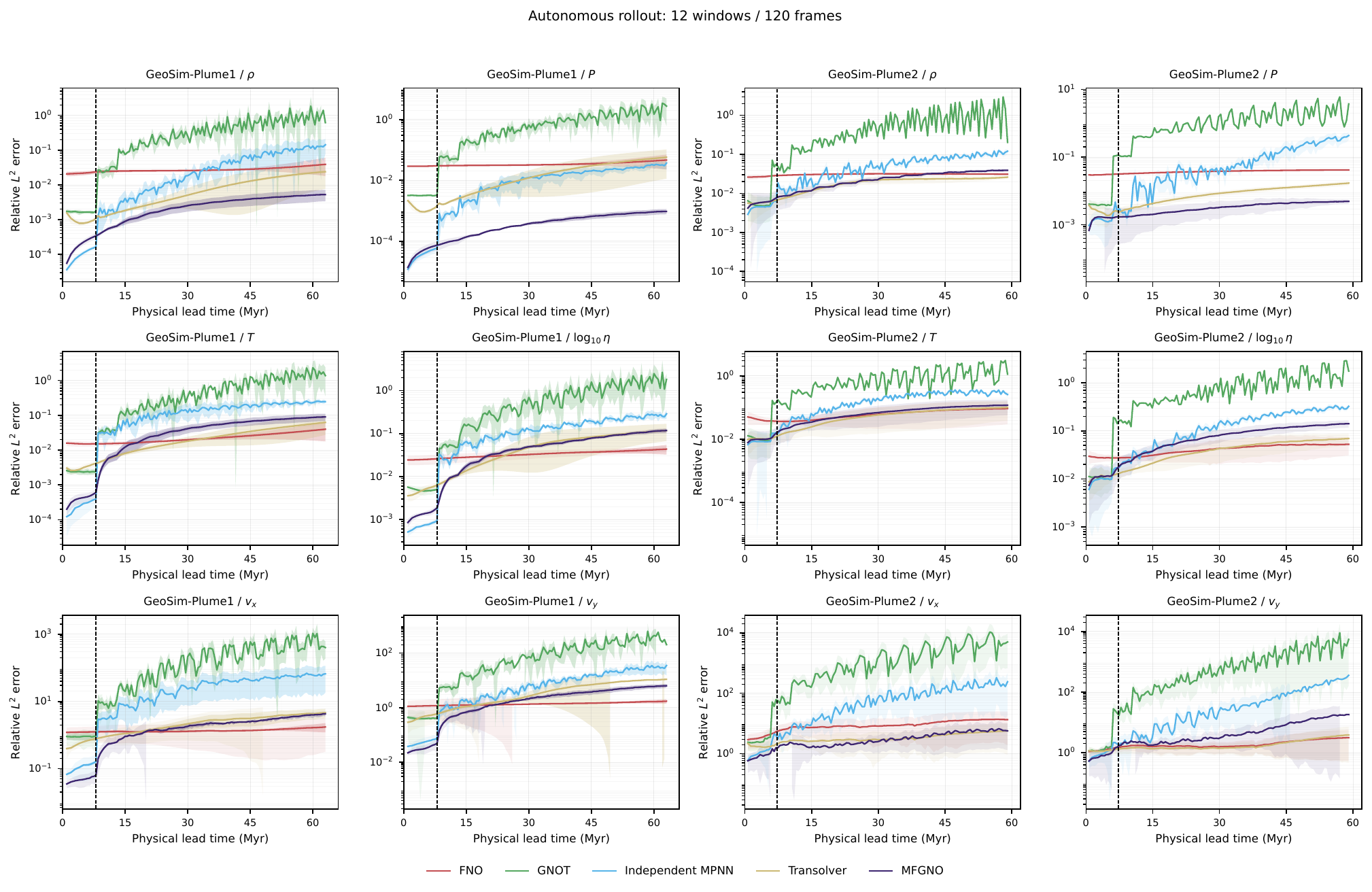}
\caption{Autonomous rollout error for all six fields on \pone\ and
\ptwo. Each panel corresponds to one physical field in one subset and
compares all evaluated models. Ground-truth fields are not reintroduced
after initialization. Errors are plotted against physical lead time
because stored snapshot intervals are non-uniform and differ across
trajectories. Relative-$L^2$ errors are shown with field-specific vertical
ranges; y-axis limits are therefore not shared across panels. For \pone,
the light band denotes trajectory-level variability (mean $\pm$ SD across
trajectories) for the first training seed and the dark band denotes
variability across the three training seeds (mean $\pm$ SD across seeds);
\ptwo\ is single-seed and therefore has no across-seed band.}
\label{fig:rollout_full}
\end{figure}

\paragraph{Failure criterion.}
Under the prespecified criterion (for each model and channel, the
$99$th percentile of frame-level relative-$L^2$ validation errors pooled
over that model's validation trajectories and frames, held for three
consecutive output windows), \model\ on \pone\ in
autonomous mode has $199$ of its $216$ trajectory--field pairs over the
threshold, $15$ outside the declared support range, and two clean pairs
($12$ trajectories $\times$ $3$ seeds $\times$ $6$ fields). FNO has
$72$ pairs clean, $113$ over threshold and $31$ outside the declared
support range. No model satisfies the criterion on all trajectory--field
pairs. These are protocol
violations, not numerical blow-up: no prediction in the evaluation became
non-finite. At a relative-$L^2$ threshold of $10^{-1}$, \model\ remains below
the threshold through $10$\,Myr on \pone; no evaluated model satisfies
this requirement over all twelve rollout windows. Fig.~\ref{fig:rollout_full}
reports the complete six-field curves.

\section{Release Manifest and Provenance}
\label{app:release}

The dataset is available at
\url{https://huggingface.co/datasets/irenier/GeoSim}. It contains the field
files and a field and coordinate dictionary; the per-trajectory parameter
metadata and fixed splits are provided with the accompanying code package.
The code package provides preprocessing scripts, model and training
configurations, evaluation code, and trained weights, and will be archived
with a persistent identifier upon publication.

\paragraph{Dataset release.}
The field data are available under CC-BY-4.0.
The dataset contains eight HDF5 files: for each subset, a
train and a test file at $128\times128$ resolution over the complete time
extent, and a train and a test file on the native $401\times309$ solver
grid in raw physical units. Validation trajectories remain within the
corresponding training-pool file and are identified by the fixed split
metadata, not stored in separate HDF5 files. Both grids describe
the same simulations; they
differ in scaling, coordinate unit, and time unit, and the dataset card
states these conversions explicitly. The
$128\times128$ fields are max-normalized by per-subset constants that are
listed on the card; the native fields are not. The native coordinates are
in meters on the non-uniform solver grid (horizontal spacing $1$--$4$ km,
vertical $1$--$22$ km), whereas the downsampled coordinates are equal-area
cell boundaries on $[0,1]$. Time is in Myr on the downsampled files and in
years on the native files. We keep each grid's native convention, not
harmonized values, and the card states the factor explicitly.
A machine-readable table of per-trajectory parameters
(\texttt{trajectory\_parameters.csv}) is provided with the code package and
carries the varying parameters, the split assignment, the row index of each
trajectory within its split file, and its physical end time; the row index
identifies the same simulation in either grid. Without that table a row of
an HDF5 file cannot be tied to a parameter draw, so it is part of the
release, not an auxiliary file. The card additionally specifies two
conventions required for correct interpretation: the pressure channel is \emph{total}
pressure, not a dynamic anomaly, and the time axis is per-trajectory and
non-uniform, so it must be read per sample, not assumed constant.

\paragraph{Release scope and exclusions.}
The continuous solver is third-party software that we neither authored nor
redistributed. Its sources, binaries and case configuration files are
therefore absent from the release. The released package ships the field
data and parameter values, not a runnable simulation. Because the
third-party solver is not redistributed, the released package does not by
itself support regeneration of the original simulations. It supports
reproduction of the sampling design, preprocessing, benchmark splits,
model training, and evaluation from the released trajectories.

\paragraph{Reproducibility of reported results.}
The accompanying code package provides scripts and machine-readable result
files for reproducing the reported dataset statistics, quantitative tables,
and evaluation figures. Per-run evaluation records store the model
configuration, seed, checkpoint, dataset split, and field-wise metrics used
to construct the reported results, so each reported metric can be traced to
its generating run. Dataset parameter metadata can be independently
re-derived from the sampling design and checked against the released
trajectory manifest. The code and data are licensed separately. The
dataset is available under CC-BY-4.0.

\end{document}